\documentclass[10pt,letterpaper]{article}
\usepackage[top=0.85in,left=2.75in,footskip=0.75in]{geometry}

\usepackage{amsmath,amssymb}

\usepackage{changepage}

\usepackage[utf8x]{inputenc}

\usepackage{textcomp,marvosym}

\usepackage{cite}

\usepackage{nameref,hyperref}

\usepackage{microtype}
\DisableLigatures[f]{encoding = *, family = * }

\usepackage[table]{xcolor}

\usepackage{array}

\newcolumntype{+}{!{\vrule width 2pt}}

\newlength\savedwidth

\newcommand\thickhline{\noalign{\global\savedwidth\arrayrulewidth\global\arrayrulewidth 2pt}%
\hline
\noalign{\global\arrayrulewidth\savedwidth}}

\usepackage{setspace} 
\raggedright
\usepackage[aboveskip=1pt,labelfont=bf,labelsep=period,justification=raggedright,singlelinecheck=off]{caption}

\makeatletter
\renewcommand{\@biblabel}[1]{\quad#1.}
\makeatother

\usepackage{lastpage,fancyhdr,graphicx}
\usepackage{epstopdf}
\fancyheadoffset[L]{2.25in}
\fancyfootoffset[L]{2.25in}
\begin{document}
\vspace*{0.2in}

\begin{flushleft}
{\Large
\textbf\newline{{\color{black}Cooperation of dual modes of cell motility promotes epithelial stress relaxation to accelerate wound healing}} 
}
\newline
\\
Michael F. Staddon\textsuperscript{1,2},
Dapeng Bi\textsuperscript{3},
A. Pasha Tabatabai\textsuperscript{4,5},
Visar Ajeti\textsuperscript{4,5},
Michael P. Murrell\textsuperscript{4,5,6},
Shiladitya Banerjee\textsuperscript{1,2*},
\\
\bigskip
\textbf{1} Department of Physics and Astronomy, University College London, London, UK
\\
\textbf{2} Institute for the Physics of Living Systems, University College London, London, UK
\\
\textbf{3} Department of Physics, Northeastern University, Boston, USA
\\
\textbf{4} Department of Biomedical Engineering, Yale University, New Haven, USA
\\
\textbf{5} Systems Biology Institute, Yale University, New Haven, USA
\\
\textbf{6} Department of Physics, Yale University, New Haven, USA
\\
\bigskip

%
%





* Correspondence: shiladitya.banerjee@ucl.ac.uk

\end{flushleft}
\section*{Abstract}
Collective cell migration in cohesive units is vital for tissue morphogenesis, wound repair, and immune response. {\color{black}While} the fundamental driving forces for collective cell motion stem from contractile and protrusive activities of individual cells, it remains unknown how their balance is optimized to maintain tissue cohesiveness and the fluidity for motion. Here we present a cell-based computational model for collective cell migration during wound healing that incorporates mechanochemical coupling of cell motion and adhesion {\color{black}kinetics} with stochastic transformation of active motility forces. We show that a balance of protrusive motility and actomyosin contractility is optimized for accelerating the rate of wound repair, {\color{black}which is} robust to variations in cell and substrate mechanical properties. {\color{black}This balance underlies rapid collective cell motion during wound healing, resulting} from a tradeoff between tension mediated collective cell guidance and active stress relaxation {\color{black}in the tissue}. 

\section*{Author summary}
Many developmental processes involve collective {\color{black}cell} motion, driven by migratory behaviours of individual cells and their interactions with the extracellular environment. 
An outstanding question is how cells regulate their internal driving forces to maintain tissue cohesiveness while promoting {\color{black}the requisite} fluidity for collective motion. Progress has been limited by the lack of an integrative framework that couples cellular physical behavior with stochastic biochemical dynamics underlying cell motion and adhesion. Here we develop a cell-based computational model for collective cell migration during epithelial wound repair that integrates {\color{black}tissue mechanics with active cell motility, cell-substrate adhesions, and actomyosin dynamics.} Using this model we show that an optimum balance of protrusive cell crawling and actomyosin contractility drives rapid directed motion of cohesive cell groups, robust to variations in cell and substrate physical properties. We {\color{black}further} show that disparate modes of individual cell migration can cooperate to accelerate collective cell migration by fluidizing confluent tissues.


\section*{Introduction}
Collective {\color{black}cell} migration is central to tissue morphogenesis, wound repair and cancer metastasis~\cite{friedl2009}. During tissue repair after wounding~\cite{grose2002}, or during closure of epithelial gaps~\cite{jacinto2002dynamic,rosenblatt2001epithelial}, collective cell migration enables the regeneration of a functional tissue. {\color{black}Gap closure} is usually mediated by two distinct mechanisms for collective cell movement~\cite{begnaud2016,martin1992actin,crawling_closure}. First, cells both proximal and distal to the gap can {\it crawl} by Arp2/3 driven forward lamellipodial protrusions~\cite{martin1992actin,crawling_closure,fenteany2000}. Secondly, cells around the gap can collectively assemble a supracellular actomyosin cable, known as a {\it purse-string}, which closes tissue voids via active contractile forces \cite{martin1992actin,bement1993}. It remains poorly understood how these two modes of collective cell movement, driven by the assembly of distinct actin network architectures, are regulated in diverse biophysical conditions.

{\color{black}Many} experimental studies have provided key insights into the physical forces driving collective cell migration~\cite{bement1993,fenteany2000,trepat2009,tambe2011,purse_string_closure,border_forces,crawling_closure}. Recent {\it in vitro} wound healing experiments have shown that closure of large wounds is initiated by cell crawling, followed by the assembly of purse string that dominates closure at smaller wound sizes \cite{purse_string_closure,border_forces}. Purse-string acts like a cable under contractile tension, pulling in the wound edge at a speed proportional to its local curvature~\cite{ladoux2015geometry}. By contrast, crawling driven closure occurs at a constant speed, regardless of wound morphology~\cite{crawling_closure}. 
However, it remains unknown how the mechanochemical properties of individual cells and their interactions with the extracellular matrix regulate crawling and purse-string based collective cell motion. 
{\color{black}While} experiments are limited in the extent to which mechanical effects are separated from biochemical processes, theoretical and computational models can decouple these variables {\color{black}precisely}.

Extensive theoretical work has been done to model collective cell migration during tissue morphogenesis and repair~\cite{sherratt1992,lee2011,banerjee2015,nagai2009computer,salm2012,basan2013,mirams2013}. {\color{black}However,} existing models do not explain how individual cells adapt their migratory machineries {\color{black}and interactions with neighboring cells} to move collectively like a viscous fluid while maintaining tissue cohesion. Continuum models of tissues~\cite{banerjee2018} as viscoelastic fluids~\cite{lee2011,border_forces} or solids~\cite{sherratt1992,kopf2013,banerjee2015,ladoux2015geometry} have been successful in {\color{black}describing} collective flow and traction force patterns observed experimentally. However, such macroscopic models cannot capture cellular scale dynamics, and therefore unsuited for connecting individual cell properties to collective cell dynamics. On the other hand, cell-based computational models, including the Cellular Potts Model~\cite{graner1992,Albert2016}, Vertex Model~\cite{honda1980,farhadifar2007}, phase-field~\cite{lober2015} or particle-based models~\cite{basan2013,tarle2015,zimmermann2016} explicitly account for dynamic mechanical properties of individual cells and their physical interactions. However, these models have not yet been developed to integrate the mechanics of cell motion with {\color{black}cell-substrate adhesions and} intracellular cytoskeletal dynamics. It remains poorly understood how migrating cells sense changes in their physical environment and translate those cues into biomechanical activities in order to facilitate collective motion. This is particularly important for epithelial wound healing, where wound edge cells actively remodel their cytoskeletal machineries {\color{black}and the resulting modes of motility} in response to changes in wound size, shapes and substrate properties~\cite{brugues2014,ladoux2015geometry,purse_string_closure}.

To overcome these limitations, we propose an integrative modeling framework that incorporates the mechano-chemical coupling of cell motion and adhesion with stochastic transformation between protrusive and contractile cell behaviors. In contrast to previous cell-based models of wound healing~\cite{nagai_vm,nagai2009computer,brugues2014}, our approach explicitly accounts for the spatiotemporal regulation of protrusive and contractile activities, cell-matrix interactions, adhesion turnover, and cell polarity. Using this model, we ask: How do migrating cells sense changes in their physical environment? How do cells regulate their modes of motilities to optimize the speed of collective motion? What roles do tissue mechanical properties play in stress propagation and relaxation during wound repair? In particular, we find that an optimum mixture of protrusive and contractile cell activities at the wound edge accelerates the rate of wound healing under diverse conditions. The optimum mixed mode of migration is robust to changes in substrate rigidity, wound shape, intercellular adhesions and cortical tension. A unique insight offered by our study is that a mixture of protrusive and contractile activities promotes faster wound repair by optimizing the tradeoff between collective cell guidance and local stress relaxation. {\color{black}Finally,} we propose {\color{black}a fundamental} mechanism by which tissues can locally fluidize to drive rapid collective cell motion while maintaining their overall mechanical integrity.

\section*{Cell-based mechanochemical model}

Our model consists of several computational components that simulate: (1) mechanical interactions between cells, (2) biochemical dynamics (protrusions, adhesions), and (3) transitions between distinct cell motility modes. Mechanical interactions between cells are simulated using the vertex model for epithelial mechanics~\cite{honda1980,nagai2009computer,farhadifar2007,mirams2013,fletcher2014,manning2010,vm_shapes}, where the geometry of each cell is defined by a two-dimensional polygon, with mechanical energy given by:
\begin{equation}\label{eq:vm}
E_i = K(A_i-A_0)^2 + \Gamma P_i^2 + \gamma P_i\;.
\end{equation}
The first term in \eqref{eq:vm} represents the energy cost for cell compressibility, where $A_i$ is the area of cell $i$, $A_0$ is the preferred cell area, and $K$ is the elastic constant. The second term, $\Gamma P_i^2$, is the energy due to contractile forces in the actomyosin cortex. The last term in \eqref{eq:vm} represents the interfacial tension between cells, which is the difference between cortical tension and the cell-cell adhesion energy per unit length. The elastic substrate is modeled as a triangular mesh of harmonic springs (Methods). Focal adhesion complexes are modeled as stiff springs that anchor the cell vertices to the substrate mesh, with attachment and detachment rates given by $k_\text{on}$ and $k_\text{off}$, respectively (see Methods). The net mechanical force acting on the cell vertex $\alpha$ is given by $ \mathbf{F}^\alpha = - \partial E_\text{tot}/\partial \mathbf{x}^\alpha$, where $E_\text{tot} = \sum_{i=1}^n E_{i} + E_\text{adh}$ is the total mechanical energy of the cells and the cell-substrate adhesions. 

\begin{figure}[htp]
\includegraphics[width=\columnwidth]{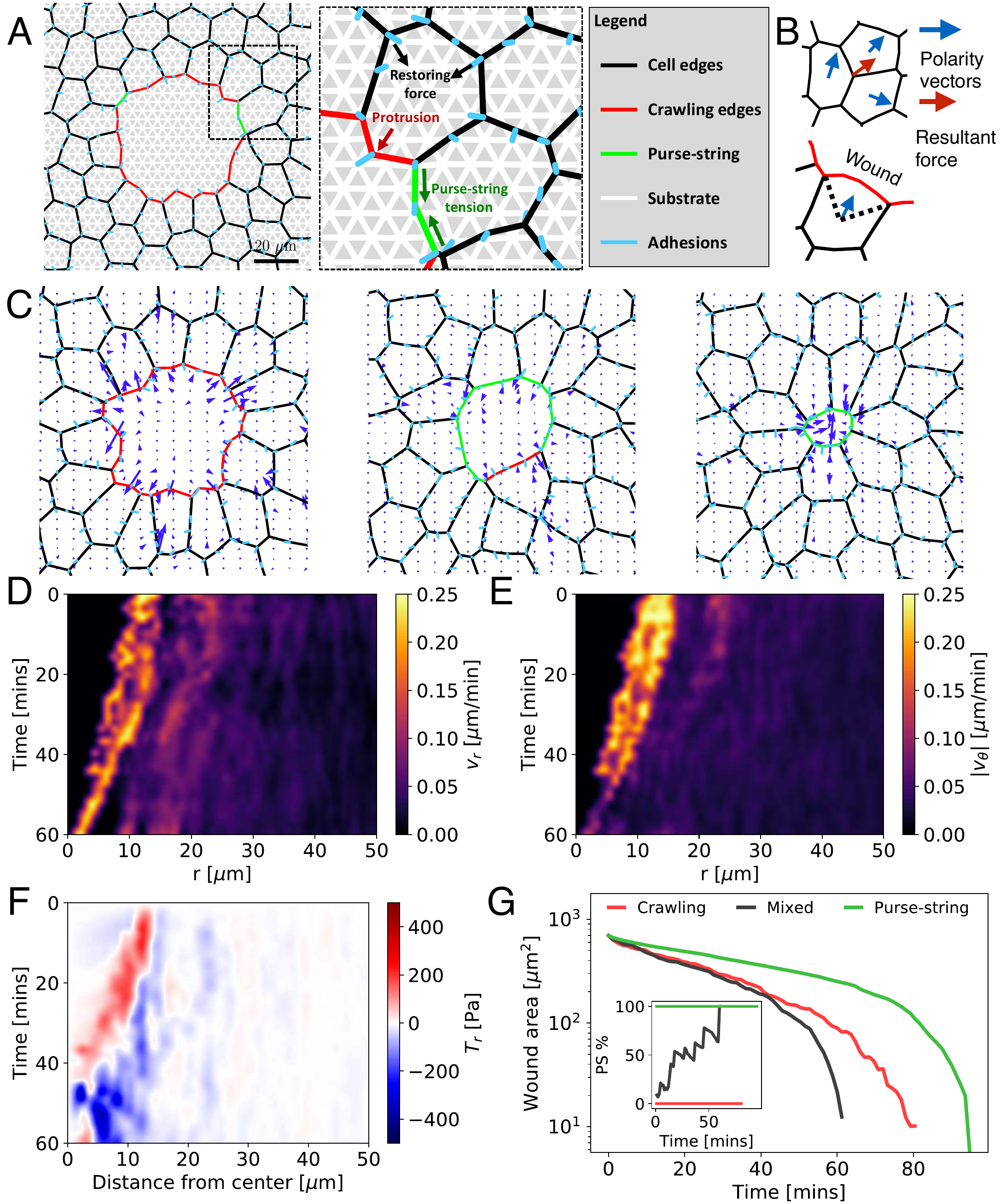}
\caption{{\bf Collective migration during wound healing driven by a mixture of crawling and purse-string based cell motility.}
A: Model schematic, showing physical forces and model elements. B: Illustration of self-propulsion force. The central vertex for a cell inside the tissue has a resultant force (red arrow) equal to the average force from its adjacent cells (blue arrows). The polarity vector (blue arrow) for a wound edge cell bisects the angle between the lines from the cell centroid to the boundary vertices. C: Tissue configuration during wound closure with $k_\text{p}=4$ hr$^{-1}$, at $t=15$ min (left), $t=33$ min (middle), $t=64$ min (right). Arrows indicate traction forces.  D-F: Kymograph of the (D) radial component of cell velocity field, $v_r$, (E) magnitude of the azimuthal velocity, $|v_\theta|$, and (F) radial traction stress, $T_r$, for the mixed modality of closure corresponding to (C). G: Log-linear plot for wound area vs time for crawling ($k_\text{p}=0$), purse-string ($k_\text{p}=1000$ hr$^{-1}$), and mixed ($k_\text{p}=4$ hr$^{-1}$) modes of closure. Inset: Time evolution of the percentage of wound perimeter covered by purse-string. See Table~\ref{table1} for the full list of default model parameters.}
\label{fig:1}
\end{figure}

In addition to mechanical forces (Fig~\ref{fig:1}A), cells within the {\color{black}bulk} tissue actively move with a self-propulsion velocity $v_0 \hat{\mathbf{p}}_i$ (Fig~\ref{fig:1}B), where $\hat{\mathbf{p}}_i$ defines the random polarity vector for cell motion, and $v_0$ is the self-propulsion speed. Cells at the wound leading edge initiate motion by crawling towards the wound center~\cite{purse_string_closure,border_forces}, with a force $\mathbf{f}_{p}$ (Fig~\ref{fig:1}A). At each time step, crawling cell fronts can transition to a purse-string at a {\color{black}constant} rate $k_\text{p}$. This leads to an increased line tension on the wound edge due to actomyosin contractility (Fig~\ref{fig:1}A) (see Methods). {\color{black}Assuming over-damped dynamics, cell vertex $\alpha$ at the wound edge moves as:
\begin{equation}\label{eq:cell}
\mu \frac{d\mathbf{x}^\alpha}{d t}= \mathbf{F}^\alpha + \mathbf{f}_p^\alpha \;,
\end{equation}
where $\mu$ is the friction coefficient. Cell vertices in the bulk of the tissue move according to following equation of motion
\begin{equation}\label{eq:cell2}
\mu \frac{d\mathbf{x}^\alpha}{d t}= \mathbf{F}^\alpha + \frac{1}{n^\alpha}\sum_{i \in \alpha} \mu v_0 \hat{\mathbf{p}}_i\;,
\end{equation}}
where the last term is the averaged self-propulsion force over $n_\alpha$ neighboring cells sharing the vertex $\alpha$ (Fig~\ref{fig:1}B). We estimate the model parameters from available experimental data (Methods, Table~\ref{table1}).

\section*{Results}
\subsection*{Cooperation of distinct modes of cell migration during wound repair}
To elucidate the mechanisms of collective cell motion during wound repair, we simulated healing of a circular wound for a mixed modality of closure: $k_p=4$ hr$^{-1}$. Initially, cells close the wound by crawling (Fig~\ref{fig:1}C), but over time they switch to the purse-string mode, resulting in rapid contraction of cell edges lining the wound periphery (Fig~\ref{fig:1}C, \nameref{S1_Video}). To quantify the spatiotemporal patterns of collective cell motion, we calculated spatially averaged radial and azimuthal velocities as a function of the radial distance from the wound center at each time point (Fig~\ref{fig:1}D-E). Initially, both radial and azimuthal velocities are highest around the wound edge and decay with distance inside the monolayer. {\color{black}As} crawling cells pull on the substrate, the resultant traction forces point radially outwards and away from the wound (Fig~\ref{fig:1}C,F). Halfway through the closure process, the purse-string fully assembles (Fig~\ref{fig:1}G-inset) and the traction forces switch to pointing radially inwards (Fig \ref{fig:1}F), in quantitative agreement with experimental data~\cite{brugues2014}. {\color{black} Consistent with experiments, tangential traction stresses are comparable in magnitude with the radial components of the traction stress (\nameref{S2_Fig}:B). Our model reproduces the experimental observation that focal adhesions are oriented towards the wound center for crawling cells~\cite{brugues2014,ajeti2018} (\nameref{S3_Fig}:A,C). By contrast, purse-string adhesions have a higher probability of orienting tangentially at the leading edge than crawling cells (\nameref{S3_Fig}:B,D,E).} As closure proceeds, the band of high radial velocities around the wound narrows (Fig~\ref{fig:1}D), while the azimuthal velocity {\color{black}narrows and} decreases around the wound (Fig~\ref{fig:1}E). This results in more coordinated inward motion of the cells. 

Increasing $k_p$ from $0$ (crawling only) to $1000$ hr$^{-1}$, monotonically increases the proportion of wound perimeter covered by the purse-string over time (Fig~\ref{fig:1}G-inset). For non-zero values of $k_p$, wound area shrinks in a biphasic manner: an initial slow exponential decay, followed by fast exponential decay, consistent with experimental data~\cite{ajeti2018}. In contrast to the mixed mode of closure (Fig~\ref{fig:1}C,G), the traction forces for crawling mediated closure are always directed radially outwards (\nameref{S2_Fig}:C), because crawling cells pull on the substrate. While further inside the monolayer the traction forces point radially inwards as the rear end of crawling cells retract via cortical contraction. In purse-string mediated closure, the wound shape remains circular throughout (\nameref{S2_Video}), in contrast to the ruffling morphology observed for crawling cell fronts (\nameref{S2_Video}). Traction forces point into the gap, and increases in magnitude as the wound size gets smaller (\nameref{S2_Fig}:D). For a fixed set of parameters, we find that a balance of purse-string and crawling mediated closure results in faster wound healing (Fig \ref{fig:1}G). To determine how the relative proportion of purse-string and lamellipodia is optimized for rapid collective motion, we turned to examine how the purse-string assembly rate ($k_p$) regulates wound closure time for varying physical properties of the cells, the underlying substrate, wound size and shape.

\subsection*{Mixture of crawling and purse-string based motilities accelerates {\color{black}wound closure} independent of substrate rigidity}
\begin{figure}[htp]
\includegraphics[width=\columnwidth]{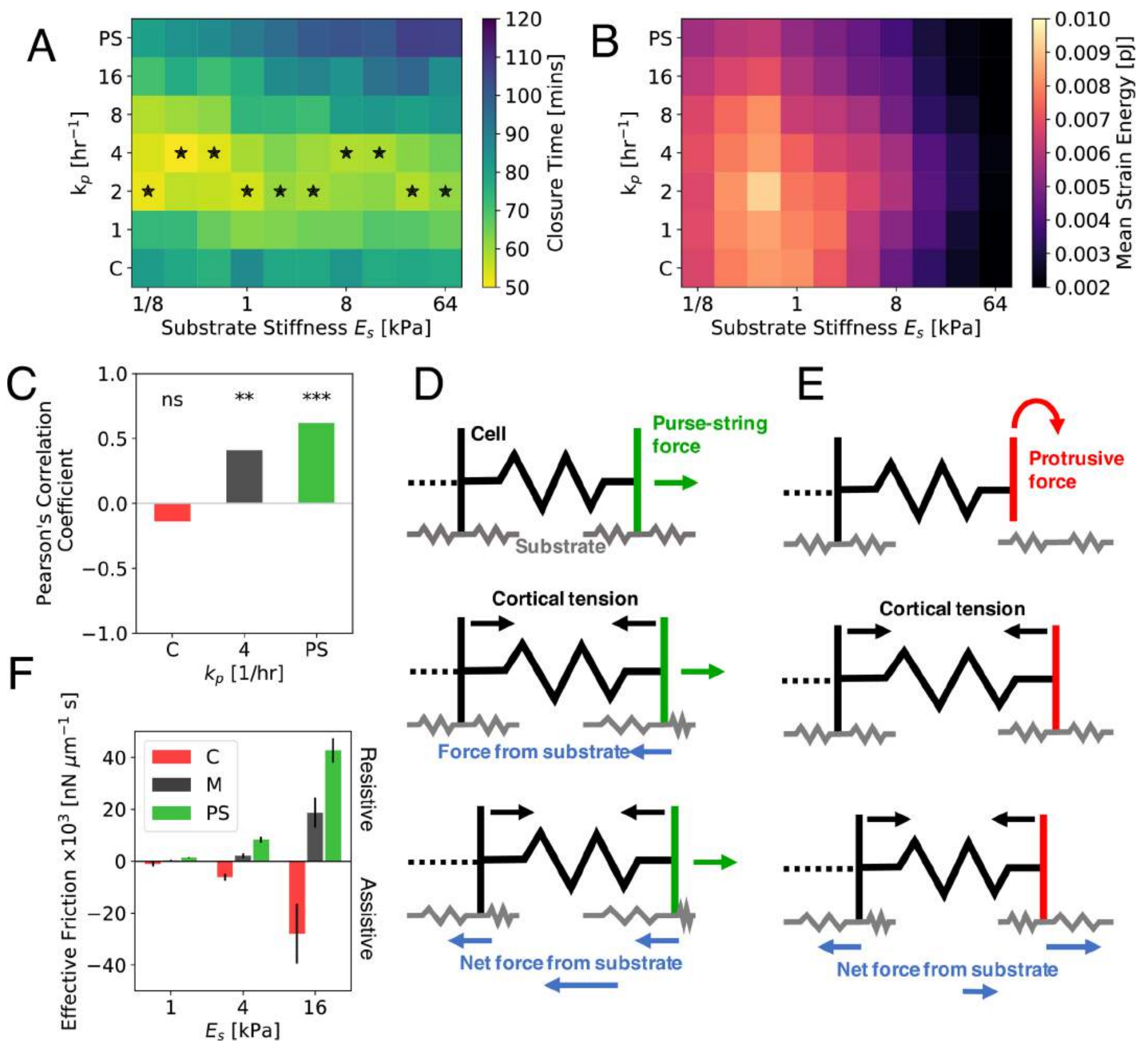}
	\caption{{\bf Mechanosensitivity of {\color{black}wound closure}.} A-B: Dependence of closure time (A) and mean strain energy (B) for different values of substrate stiffness ($E_s$) and $k_\text{p}$. Starred cells indicate the fastest wound closure for a given $E_s$ with varying $k_p$ (PS: $k_\text{p}=1000$ hr$^{-1}$; C: $k_p=0$). Each data point corresponds to the average of 5 simulations. C: Pearson's correlation coefficient, $r$, between closure time and substrate stiffness, for crawling ($k_\text{p}=0$), purse-string ($k_\text{p}=1000$ hr$^{-1}$), and mixed ($k_\text{p}=4$ hr$^{-1}$) modes of closure. The asterisk represents coefficients significantly different from zero (p$< 0.05$); ** means p$< 0.01$, and *** means p$< 0.001$. For each mode, n = 25. D-E: Schematic of purse-string driven (D) and crawling mediated (E) cell motility on an elastic substrate. Blue arrows represent {\color{black}reaction force} from the substrate, green (red) arrows represent purse-string (protrusion) driving forces, and black arrows represent cortical contractile forces. F: Effective friction from the substrate on the leading edge of the wound for crawling ($k_\text{p}=0$), purse-string ($k_\text{p}=1000$ hr$^{-1}$), and mixed ($k_\text{p}=4$ hr$^{-1}$) modes of migration. Error bars represent standard deviation (n = 5).}
	\label{fig:3}
\end{figure}

Since the speed of cell crawling and the magnitude of traction forces are sensitive to substrate rigidity~\cite{discher_stiffness,schwarz2013}, we first investigated the role of substrate stiffness on {\color{black}wound closure time}. To this end, we varied the substrate Young's modulus, $E_s$, and the purse-string assembly rate, $k_\text{p}$, for fixed physical properties of the tissue and the wound. We find that wound closure time increases with $E_s$ for higher values of $k_\text{p}$, {\color{black}but remained insensitive for crawling mediated closure} (Fig~\ref{fig:3}A). {\color{black}Strikingly}, there exists an optimum value of $k_\text{p}$ (corresponding to mixed modality) for any value of $E_s$, which results in minimum closure time (Fig~\ref{fig:3}A). For fixed $k_\text{p}$, strain energy transmitted to substrate decreases monotonically with increasing {\color{black}stiffness for $E_s>0.5$kPa} (Fig~\ref{fig:3}B) (see Methods for calculation details). For all values of $E_s$ and $k_p$, faster {\color{black}wound closure} coincides with higher strain {\color{black}energy} transmitted to the substrate, signifying a {\color{black}positive correlation between energy cost and the speed of} wound healing. 

{\color{black}Our results agree with experimental findings that wound closure time is not sensitive to changes in substrate stiffness for moderate to high rigidities~\cite{brugues2014,ajeti2018}. On very soft substrates ($<500$ Pa), our model predictions are inconsistent with experiments by Anon et al~\cite{crawling_closure}, who showed that crawling-based migration fails to close wounds on very soft gels ($\sim 100$ kPa), as lamellipodia do not form. This may be captured by implementing additional biochemical feedback mechanisms between protrusive activity and substrate stiffness, beyond the scope of our mechanical model.}


As $E_s$ is increased, purse-string driven motion slows down. To quantify the dependence of closure time on stiffness, we calculated the Pearson's correlation coefficient between wound closure time and substrate stiffness for {\color{black}different modes of wound closure} (Fig~\ref{fig:3}C). We find that purse-string based motility slows down with increasing stiffness, with a positive correlation coefficient significantly different from zero (p-value $< 0.05$). In contrast, crawling based motility and have the least significant correlation coefficient (p-value $> 0.05$). 

The sensitivity of purse-string driven motility to substrate rigidity (Fig~\ref{fig:3}A-C) can be explained by a mechanical force balance argument (Figs.~\ref{fig:3}D-E). Purse-string driven contractile forces drag the border cells into the gap, in competition with cortical tension retracting the rear cell edges. This results in a large net resistive force from the deforming elastic substrate (Fig~\ref{fig:3}D). By contrast, crawling cells pull the substrate backwards at the wound edge and contractile forces pull the substrate forward at the cell rear (Fig~\ref{fig:3}E). This dipole-like traction pattern results in {a net assistive force from the substrate}, pointing towards direction of cell crawling. During mixed mode of migration, a combination of net {\color{black}assistive} and resistive forces {\color{black}should therefore lead to} the least sensitivity to substrate stiffness.

To test this hypothesis, we {\color{black}computed} the net radial traction force, $F_r$, on the substrate under the first row of cells at the wound edge. We then calculate the time-averaged ratio between the radial force and the radial velocity, $v_r$, of the wound edge, to obtain an effective friction coefficient: {\color{black}$\mu_\text{eff}=\langle F_r/v_r \rangle$} (Fig~\ref{fig:3}F). We find that {\color{black}$\mu_\text{eff}$ monotonically increases in magnitude with increasing substrate stiffness} (for all modes of migration), consistent with previous theoretical predictions~\cite{walcott2010}. For all values of substrate stiffness, purse-string motion leads to the highest {\color{black}positive $\mu_\text{eff}$}, suggesting high resistance and sensitivity to substrate rigidity. Crawling driven motility {\color{black}leads to negative $\mu_\text{eff}$, indicative of assistive motion. By contrast,} the mixed mode of migration leads to the lowest {\color{black}magnitude of $\mu_\text{eff}$, i.e. least drag from the substrate}.

{\color{black}Rigidity sensing by different modes of collective migration is expected to be strongly coupled to focal adhesion kinetics. While we have assumed constant rates of binding and unbinding of cell-substrate adhesions, experiments have demonstrated that integrin-ligand pairs form catch bonds~\cite{kong2009}, such that $k_\text{off}$ decreases under low forces and increases under larger forces. To test the if the mechanosensitivity of cell-substrate adhesion bonds impact our results, we implemented a catch bond model for adhesions, assuming a single bound state and two unbinding pathways~\cite{pereverzev2005} (see Methods). As a result, the crawling mode of closure is now more sensitive to changes in substrate stiffness, with closure time increasing with stiffness, before decreasing at higher stiffnesses due to increased adhesion lifetime (\nameref{S4_Fig}). Purse-string driven closure shows an increase in sensitivity compared to the default case, while the mixed mode of closure is least sensitive to changes in substrate stiffness. However, the mixed mode of migration is always the fastest, irrespective of force sensitivity of the adhesions.}

Aside from {\color{black}mechanosensitivity of different modes of wound closure}, the driving force for closure is expected to be strongly dependent on the relative proportion of purse-string and crawling cells. Since actomyosin purse-string is a cable under tension, the driving force for closure is proportional to the wound curvature. As a result, purse-string driven closure is expected to be sensitive to the wound geometry~\cite{ladoux2015geometry,purse_string_closure}. By contrast, crawling driven closure has been found to reduce wound area at a constant speed \cite{crawling_closure}. Therefore, we sought to investigate how the coaction of purse-string and crawling based motilities modulate collective motion for varying wound morphologies.

\subsection*{Wound geometry regulates the optimum modality of collective motion}
\begin{figure}[!h]
	\includegraphics[width=\columnwidth]{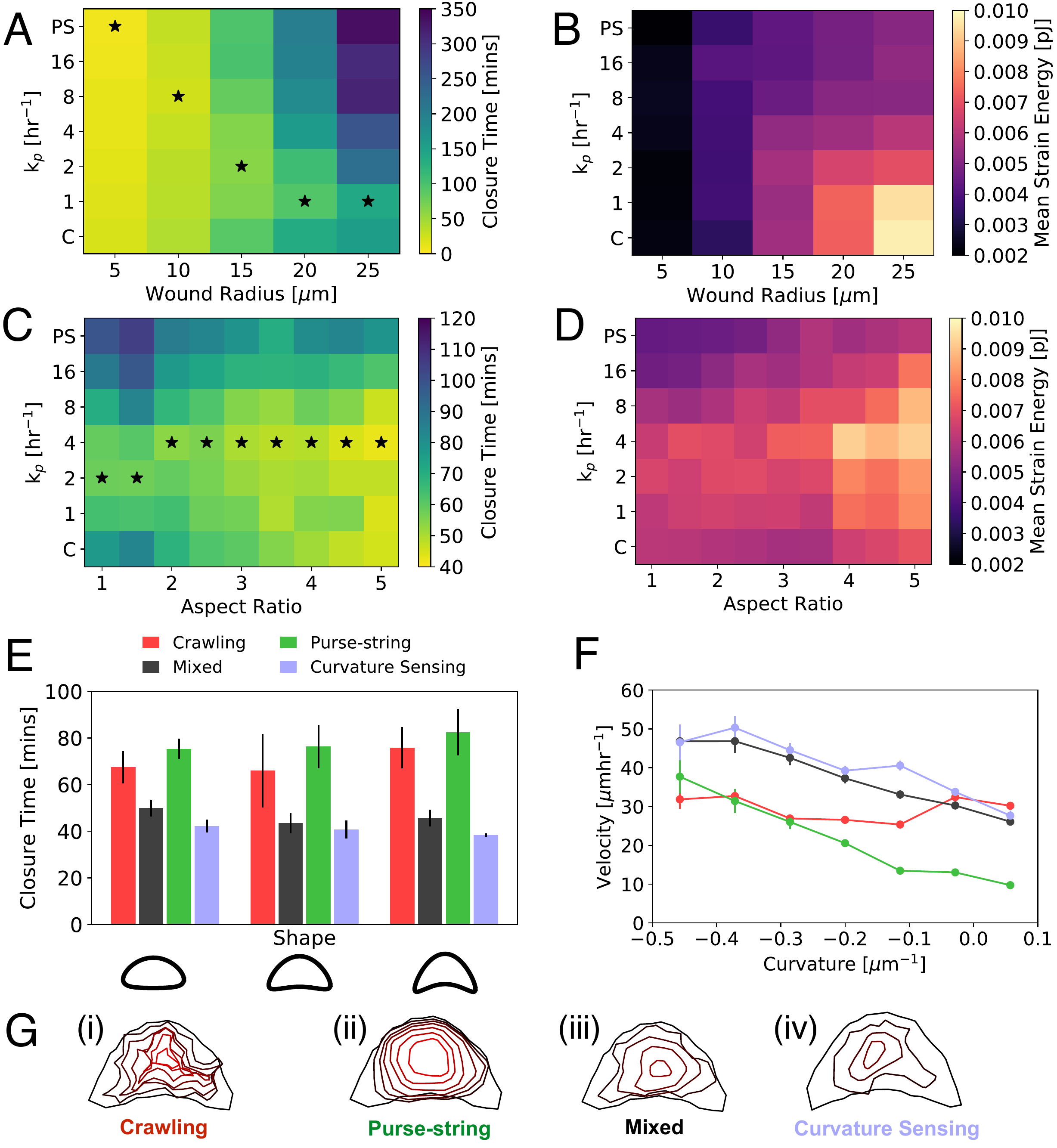}
	\caption{{\bf Wound geometry regulates the optimum modality of collective migration.} A: Closure time for different values of wound radius and $k_\text{p}$. B: Mean strain energy vs wound radius for different values of $k_p$. C: Closure time for different values of wound aspect ratio and $k_\text{p}$. D: Substrate strain energy as a function of wound aspect ratio and $k_\text{p}$. E: Wound closure time for concave wound shapes for crawling ($k_\text{p}=0$), purse-string ($k_\text{p}=1000$ hr$^{-1}$), mixed ($k_\text{p}=4$ hr$^{-1}$) {\color{black}and curvature-sensing} modes of closure. Error bars represent standard error of mean ($n=5$). F: Velocity against curvature during simulations of the right most shape in (E) for crawling, purse-string and mixed modes of closure. {\color{black}G: Evolution of wound morphology during closure by (i) pure crawling, (ii) pure purse-string, (iii) a combination of crawling and purse-string, and (iv) curvature-dependent formation of purse-string.} Colors progressively change from black to red with increasing time.}
	\label{fig:2}
\end{figure}

For circular wounds of varying radii we recapitulate the experimental result that closure time increases with wound radius (Fig~\ref{fig:2}A)~\cite{crawling_closure}. 
However, the optimum purse-string assembly rate ($k_\text{p}$) for fastest closure decreases with wound radius, such that closure time is highly sensitive to $k_\text{p}$ for larger wounds. This is because purse-string driven forces are higher near the end of closure, and that purse-string force is low in the beginning of closure of a large wound. For larger wound radii, an optimum mixture of purse-string and protrusive cell crawling leads to fastest closure.
We find that the average strain energy on the substrate increases monotonically with wound radius for $k_p$ (Fig~\ref{fig:2}B), but is more sensitive to wound size for purely crawling mediated migration ($k_p=0$). 

Next we simulated elliptical shaped wounds of fixed area but varying aspect ratios. We find that regardless of the migratory mode, closure time {\color{black}decreases} with increasing aspect ratio (Fig~\ref{fig:2}C). In addition, there exists an optimum value of $k_\text{p}$ for a given aspect ratio that leads to minimal closure time. {\color{black}Thus, a mixed mode of closure is always the fastest, but isn't much faster than crawling mediated closure for high aspect ratio wounds. This is because} crawling cells advance at a constant speed perpendicular to the wound edge. Therefore only the short axis distance must be crossed for the wound to close (\nameref{S4_Video}) (\nameref{S5_Fig}:A). For purse-string driven closure, the high curvature ends of elliptical wounds move rapidly inwards, leading to faster closure than circular shapes (\nameref{S5_Fig}:C). At all values of aspect ratio, strain energy is inversely proportional to closure time (Fig~\ref{fig:2}D).

Since purse-string behaves as a contractile cable, then for wounds with concave morphologies (positive curvatures), cells should be pulled away from the wound by the purse-string tension. To investigate this we simulated concave wound shapes as in ref. \cite{ladoux2015geometry}. For varying degrees of concavity (with fixed area), we observed that a mixed mode of closure leads to fastest wound closure (Fig~\ref{fig:2}E). To quantify the relationship between wound healing speed and curvature, we measured the local velocity and curvature at the wound perimeter. We find that the purse-string velocity is proportional to the curvature, crawling velocity is curvature-independent, while a mixture of crawling and purse-string leads to faster collective motion, with velocity decreasing with curvature (Fig~\ref{fig:2}F, \nameref{S5_Video}). These findings quantitatively agree with experimental data~\cite{ladoux2015geometry}.

{\color{black} Previous studies suggest the possibility that purse-string and lamellipodia-based migration during wound healing can be geometrically coupled~\cite{brugues2014,ladoux2015geometry}, such that the formation of protrusive borders may be directly coupled to the assembly of purse-string cables on neighboring wound edges with opposite curvatures. Such a mechanism is not captured by a purely stochastic transition between protrusive and contractile activities. To this end, we implemented a model of curvature sensing motility of the wound leading edge, where the switching between crawling and purse-string mechanisms is regulated by the local curvature of the wound (\nameref{S7_Fig}). Based on this model, if the curvature of a cell's leading edge is larger than a threshold curvature, it contracts via purse-string. Otherwise, the cell moves via protrusive crawling (Methods, \nameref{S7_Fig}). We applied this model to wounds with non-uniform curvatures as in Fig~\ref{fig:2}E. Consequently, the convex regions move forward by crawling, whereas contractile purse-string cables assemble in the concave regions. We find that for all three concave shapes in Fig~\ref{fig:2}E, the curvature sensing mechanism closes the wound at least as fast as in the mixed case with stochastic switching of motility modes (Fig~\ref{fig:2}E-G). We note that the curvature-sensing mechanism may not be applicable to the closure of undamaged epithelial gaps where purse-string cables do not form~\cite{crawling_closure}.
}
\subsection*{Optimum balance of protrusive and contractile cell activities promotes rapid wound healing via active stress relaxation}

\begin{figure}[!h]
	\includegraphics[width=\columnwidth]{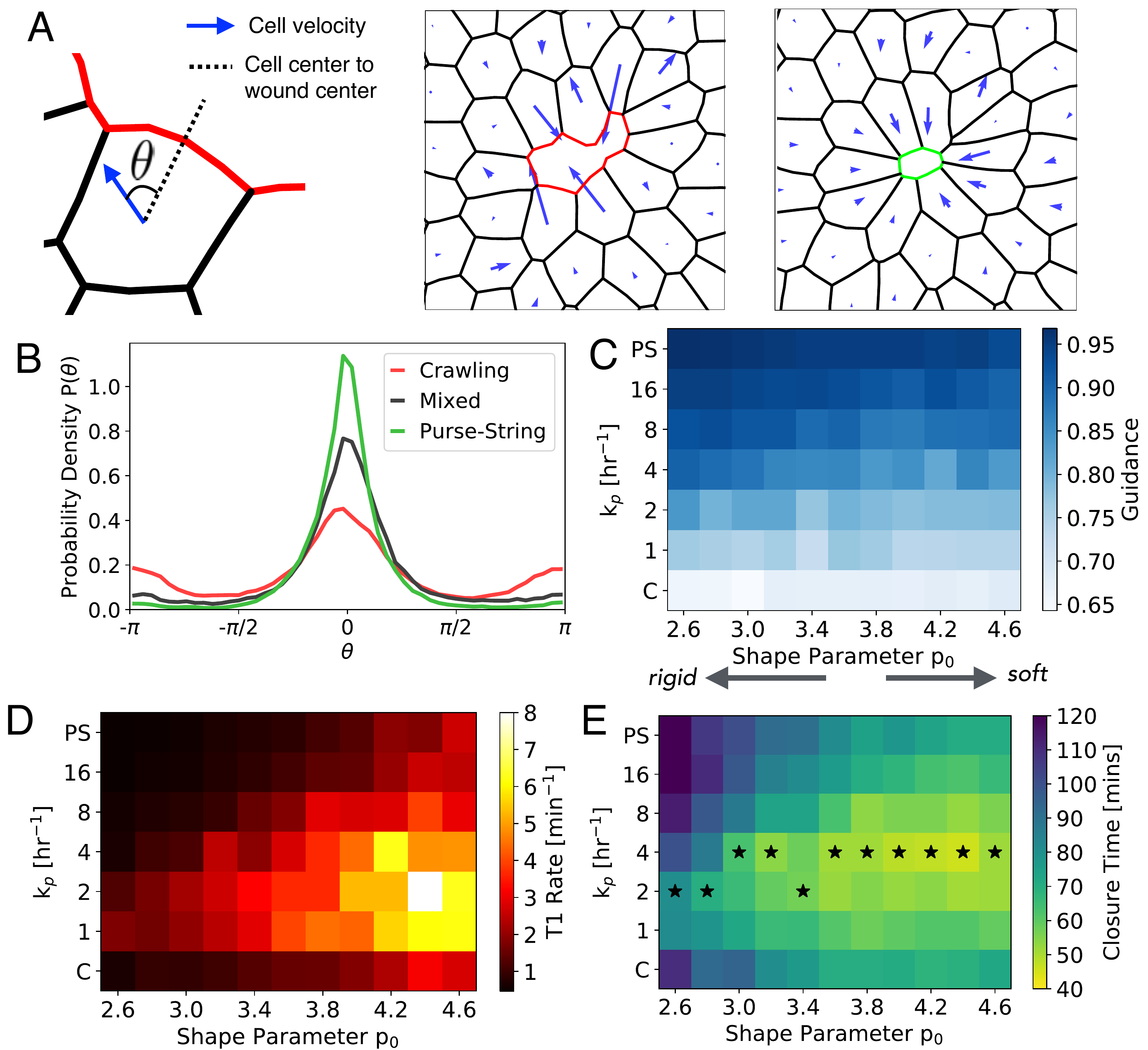}
	\caption{{\bf Tension mediated cell guidance and active stress relaxation promotes rapid wound healing.} A: Definition of the angle $\theta$ between the cell center velocity and the radial vector to wound center. Right: Representative cell velocity fields for crawling and purse-string modes of closure. B: Probability density distribution for $\theta$ for crawling ($k_\text{p}=0$), purse-string ($k_\text{p}=1000$ hr$^{-1}$), and mixed ($k_\text{p}=4$ hr$^{-1}$) modes of closure. C: Guidance parameter for different values of shape parameter $p_0$ and $k_p$. D: Rate of T1 transitions in the wounded tissue for varying $p_0$ and $k_p$. E: Closure time for different values of $p_0$ and $k_\text{p}$. Starred cells indicate the fastest wound closure for a given $p_0$.}
	\label{fig:4}
\end{figure}

Our cell-based model predicts many differences in collective cell motility driven by contractile and protrusive activities (Fig~\ref{fig:1}-\ref{fig:2}). In particular, purse-string tension rounds the wound edge and leads to solid-like, radial deformation of the tissue (Fig~\ref{fig:2}F-inset). By contrast, crawling cells ruffle the wound leading edge (Fig~\ref{fig:2}G, \nameref{S3_Video}, \nameref{S5_Video}), {\color{black}suggestive of lack of guided motion}. To quantify differences in tissue deformation and their relationship to collective motion, we measured the angle ($\theta$) between cell center velocity and the unit vector pointing towards the wound center (Fig~\ref{fig:4}A). In purse-string driven closure ($k_\text{p}=1000$ hr$^{-1}$), the angle distribution shows a single peak at $\theta=0$, corresponding to radially inward deformation (Fig~\ref{fig:4}B). By contrast, crawling cells ($k_\text{p}=0$) have a wider distribution of angles, with secondary peaks at $\theta=\pm \pi$ (Fig~\ref{fig:4}B), representing outward motion from cell neighbor exchanges (Fig~\ref{fig:4}A). To quantify the distributions, we define {\it collective cell guidance}, $\mathcal{G}$, as the probability that a cell moves towards the wound center: $\mathcal{G}$=$\int_{-\pi/2}^{\pi/2}P(\theta)d\theta$, which monotonically increases with increasing $k_p$ (Fig~\ref{fig:4}C). 

Since tissue deformation properties depend on cortical tension, cell contractility, and cell-cell adhesions~\cite{foty2005,mertz2012,mertz2013,manning2010}, we investigate how cellular mechanical properties regulate collective guidance ($\mathcal{G}$). We can rewrite the mechanical energy of cells (Eq~\eqref{eq:vm}) as:
\begin{equation}
E_i=K(A_i-A_0)^2 + \Gamma (P_i-P_0)^2\;,
\end{equation} 
where $P_0=-\gamma/2\Gamma$ is the preferred cell perimeter. The non-dimensional shape parameter $p_0=P_0/\sqrt{A_0}$ controls cell shape anisotropy and the emergent rigidity of confluent tissues~\cite{bi2015}. Increasing $p_0$ reduces cortical tension relative to cell-cell adhesions, which softens the tissue. It has been shown that confluent tissues behave like a jammed solid for $p_0<3.81$, whereas it exhibits fluid-like behaviour for $p_0>3.81$~\cite{bi2015}. Activity in the form of cell motility, division, or death can fluidize tissues further by lowering the critical $p_0$ for rigidity transition~\cite{ranft2010,bi2016,barton2017}. In our model, activity arises from self propulsion ($v_0$) (\nameref{S1_Fig}:B, \nameref{S8_Fig}), and cell crawling whose relative strength is regulated by $k_p$. We find that increasing $p_0$ decreases $\mathcal{G}$, regardless of $k_p$ (Fig~\ref{fig:4}C). The decrease in $\mathcal{G}$ with increasing $p_0$ arises from an increased rate of cellular neighbor exchanges (T1 transitions) that locally fluidizes the tissue (Fig~\ref{fig:4}D). Surprisingly, for a fixed $p_0$, T1 rates in the wounded tissue is highest for intermediate values of $k_p$, resulting in minimum closure time (Fig~\ref{fig:4}E). With higher $p_0$, cells have a higher preferred perimeter, such that both contractile and protrusive motilities experience lower mechanical resistance from tension in the border cells (\nameref{S6_Video}) (\nameref{S9_Fig}). This enables a faster reduction in wound area as compared to rigid tissues with lower $p_0$ (Fig~\ref{fig:4}E).  

These findings elucidate the mechanical basis for rapid collective migration via a mixture of protrusive and contractile cell activities. Purse-string driven tension maximizes collective cell guidance and leads to the lowest frequency of tissue rearrangements, such that cell movements are impeded by mechanical resistance from the surrounding tissue. By contrast, purely crawling motion exhibits the lowest collective guidance due to randomized protrusions of individual cells at the wound leading edge. We find that an optimum mixture of crawling and purse-string leads to intermediate collective guidance, while maximizing the frequency of local tissue rearrangements (Fig~\ref{fig:4}D). This mechanism of active fluidization enables tissues to locally relax their mechanical stress, promoting rapid wound healing. {\color{black}When intercalations are disabled in the model, tissue mechanical energy increases due to increase in cell elongation around the wound (\nameref{S10_Fig}). This results in cell jamming and slowing down of wound closure. Therefore, cell intercalations, promoted by a mixture of contractile and protrusive forces, lead to efficient wound closure by minimizing both tissue mechanical energy and wound closure time. Recent experiments, however, suggest that cells may not necessarily try to minimize energy or closure time during wound healing~\cite{ajeti2018}. But rather, they tend to coordinate the assembly of diverse actin architectures to conserve the amount of mechanical work done per unit time.} 

\section*{Discussion}

Our cell-based computational model quantitatively captures a wide range of experimental trends including the patterns of collective cell motion and traction stress organization for crawling and purse-string mediated wound closure (Fig~\ref{fig:1}). We reproduced the experimentally observed size-dependence of wound closure times, the curvature dependence of purse-string velocity, and independence of cell crawl speeds to variations in wound morphology. We predict that increasing aspect ratio of the wound speeds up closure as crawling cells can rapidly cross the short axis of the wound, whereas purse-string cables can generate rapid movements on regions of high curvature (Fig~\ref{fig:2}). 

Robust to variations in substrate and tissue mechanical properties, we find that an optimum proportion of protrusive and contractile motilities accelerates wound closure. While purse-string driven motion slows down on stiffer gels due to an increased resistance from drag on the substrate, crawling driven migration is largely independent of substrate stiffness (Fig~\ref{fig:3}). {\color{black}We find} that a mixed mode of collective migration is more efficient regardless of substrate stiffness. Robust to parameter variations, an increase in closure speed is associated with an increase in the strain energy transmitted to the underlying substrate (\nameref{S11_Fig}). As a result, migrating cells actively dissipate more mechanical energy to their environment in order to speed up collective motion.

A source of active stress dissipation comes from cellular neighbor exchanges that locally fluidize the tissue, resulting in faster wound closure (Fig~\ref{fig:4}). {\color{black}These T1 transitions have previously been observed {\it in vivo}, during wound closure in {\it Drosophila} embryo epidermis~\cite{razzell2014}. T1 transitions are also observed in our {\it in vitro} laser-ablation experiments on MDCK monolayers, where the number of cells at the wound edge decreases over time via wound edge intercalations (\nameref{S12_Fig})}. In our model, the mechanism of active fluidization via intercalation is promoted by a mixture of protrusive and contractile activities of wound edge cells, and reduced contractility or increased cell-cell adhesion in the bulk of the tissue. The ability to actively remodel an elastic tissue, coupled with tension-driven collective cell guidance, constitute the two key mechanisms for rapid directed motion in adherent environments. While the stress relaxation mechanism in our model comes only from cell neighbor exchanges, other dissipative mechanisms can also be triggered by mechanical forces including cell shape fluctuations~\cite{curran2017}, cell division~\cite{wyatt2015} or cell death~\cite{rosenblatt2001epithelial}. In these cases, our prediction will remain very similar, with the rate of cell movement into free space augmented by the sum of relaxation rates of various dissipation modes~\cite{ranft2010}. A future challenge is to identify the molecular pathways that activate distinct stress relaxation modes during tissue development and regeneration.

\section*{Methods}

\subsection*{Cell-substrate interactions}

We model the substrate as a triangular mesh of springs with a spring constant $k_s$. The Young's modulus of the substrate is given by $E_s=2k_s/\sqrt{3}h_s$, where $h_s$ is the substrate thickness, and the Poisson's ratio for a triangular mesh is $\nu = 1/3$.

Since focal adhesions and cellular traction forces typically localize at the cell periphery~\cite{mertz2013}, we implement adhesions at the cell boundaries. We model the focal adhesion complexes as stiff springs with stiffness $k_f$, which connect the cell vertices with the substrate mesh. Bound focal adhesions can detach stochastically with a rate $k_\text{off}$, whereas unbound cell vertices can attach to the nearest node of the substrate mesh with a rate $k_\text{on}$. The resultant force on the cell vertex is,
\begin{equation}
{\bf f}_{\text{adh}}^\alpha=-\frac{\partial E_\text{adh}^\alpha}{\partial \mathbf{x}^\alpha}\;,
\end{equation}
\begin{equation}
E_\text{adh}^\alpha=\sigma_\alpha\frac{k_f}{2}  (|\mathbf{x}^\alpha - \mathbf{r}^\alpha| - |\mathbf{x}_0^\alpha - \mathbf{r}_0^\alpha|)^2\;,
\end{equation}
where $E_\text{adh}^\alpha$, is the adhesion energy, $\sigma_\alpha$ is the state variable for cell-substrate attachment (0: detached; 1: attached), $\mathbf{r}^\alpha$ is the position of the substrate mesh connected to ${\bf x}^\alpha$, and $\mathbf{x}_0^\alpha$ and ${\bf r}_0^\alpha$ are the initial positions of the cell and the substrate vertices at the time of adhesion formation. 

\subsection*{Active cell motility}

Each cell carries a unit polarity vector, $\hat{{\bf p}}_i$, which represents the front/rear polarization of a motile cell~\cite{szabo2010}. {\color{black}The polarity vector is an internal state variable of cell that specifies the preferred orientation of cell motion, not their actual direction of motion}. Cells in the bulk of the tissue, i.e. not on the wound edge, move due to self-propulsion \cite{bi2016}. The polarity of a bulk cell $i$ is defined by a unit vector with angle $\theta_i$ that undergoes rotational diffusion:
\begin{equation}
\partial_t \theta_i = \eta_i(t),\ \langle \eta_i(t) \eta_j(t') \rangle = 2D_r \delta_{ij}\delta(t-t')\;,
\end{equation}
where $D_r$ is the rotational diffusion constant, and $\eta_i(t)$ is a Gaussian white noise with mean $0$ and variance $2D_r$. The self-propulsion of cell $i$ results in a force on the vertex $\alpha$ as: $\frac{1}{n_\alpha}\sum_{\alpha \in i} \mu v_0 \hat{\mathbf{p}}_i$,
where $v_0$ is the self-propulsion speed, and the sum is over all neighboring cells to vertex $\alpha$ (\nameref{S1_Fig}:B). 

{\color{black}Here, we have neglected alignment interactions between cell polarity vectors in the bulk of the tissue, which can drive coherent swirling motion of cell collectives~\cite{schaumann2018}. Without such polarity alignment rules, cell velocity vectors remain correlated over $\sim 5$ cell diameters due to mechanical interactions (\nameref{S13_Fig}), somewhat less than the correlation lengths measured in experiments in the absence of a wound~\cite{angelini2010}.}

To model lamellipodia based crawling, we allow cell vertices at the wound edge to protrude in the direction of polarity before attaching to the substrate (\nameref{S1_Fig}). This pushes the cell front outwards, while cortical tension pulls the rear of the cell forwards. The polarity vector of cells along the wound points into the gap, and is determined by the mid-point of the wound edges. {\color{black}The direction of protrusion is given by the unit vector} $\hat{\mathbf{v}}_\alpha^i$ of wound cell $i$, which makes half the angle between the two lines joining the centroid of cell $i$ to the vertices on the wound that neighbour other cells, i.e. are on the boundary of internal and external edges (\nameref{S1_Fig}:C). This ensures contact inhibition of locomotion~\cite{zimmermann2016}, preventing collision of two neighbouring cells. For a cell $i$ neighbouring the wound, the crawling force on vertex $\alpha$ on the wound edge is given by: ${\bf f}_\text{p}^\alpha= f_p(1-\sigma_\alpha) \hat{\mathbf{v}}^\alpha_i$, where $f_p$ is the protrusion force magnitude. {\color{black}For simplicity we have assumed that $f_p$ is independent cell-substrate adhesions. However, protrusive activity remains strongly correlated to focal adhesion kinetics, since the frequency of the protrusions is slaved by the rate of focal adhesion binding and unbinding. As a consequence of this feedback, increasing the duty ratio of adhesions leads to slower crawl speeds and increased closure time (\nameref{S4_Fig}:D).}
{\color{black}
\subsection*{Curvature sensing model for purse-string formation}
Here we describe the model where the switching between crawling and purse-string modes is dependent on the local geometry of the wound leading edge. At each time step in the simulation, cells at the wound edge makes a decision to switch its motility phenotype based on the local curvature of the wound edge. We calculate the curvature of a wound edge cell as the inverse of the radius of a circle inscribed to that cell edge. Curvature is defined as positive if the wound is convex (e.g. a circle), and negative otherwise. If the curvature is above a threshold value, then the cell switches to a purse-string mode. If the curvature is below the threshold value, then the cell moves by crawling. As a result, cells typically start by crawling and switch to the purse-string mode as the wound shrinks in size, consistent with experimental findings~\cite{brugues2014}.

To determine the optimum value of the threshold curvature, we varied the threshold curvature for switching to a purse-string mode, and computed the resultant wound closure time for a given initial wound shape. The optimum threshold curvature is given by the curvature value that minimizes wound closure time, as shown in \nameref{S7_Fig}:B. 

\subsection*{Catch bond model for cell-substrate adhesions}
We implemented a catch-bond model for cell-substrate adhesions, where the detachment rate of the adhesion bonds, $k_\text{off}$ is a function of the bond tension, $f$, as given below:
\begin{equation}
k_\text{off}(f)=k_0e^{-f/f_0} + k_1 e^{-f/f_1}\;.
\end{equation}
The functional form for the detachment rate is taken from a catch bond model for integrin-ligand bonds that assumes a single bound state and two unbinding pathways~\cite{pereverzev2005}. The parameters $k_0$, $k_1$, $f_0$, and $f_1$ have previously been estimated for single integrin ligand bonds~\cite{oakes2018}. Based on that estimate, we calibrate these parameters for the coarse-grained adhesion bonds in our simulations that represent several ligand-integrin pairs. We used parameter values of $k_0 = 25$ hr$^{-1}$, $k_1 = 0.006$ hr$^{-1}$, $f_0 = 3.125$ $\mu$N, and $f_1 = 0.6944$ $\mu$N, which results in the default unbinding rate at zero force, and showed high sensitivity to substrate stiffness in the range 1-16 kPa.
}
\subsection*{Model Implementation}
The vertex model is implemented using Surface Evolver \cite{surface_evolver}. We generate a wound by removing any cells that lie totally or partially within the wounded area. Edges surrounding the wound are then moved to the target wound shape. We then relax the energy of the remaining cells without adhesions so that all vertices on the wound lie on the target wound perimeter and system is at an energy minimum. To initiate gap closure, cells around the wound are set to crawling mode. We then execute the following steps (\nameref{S1_Fig}) until wound closure:
\begin{itemize}
	
	\item Update adhesion states for cell vertices. Adherent vertices attempt to unbind with a rate $k_\text{off}$ at each time step. Detached vertices attempt to attach to the nearest node of the substrate mesh with a rate {\color{black}$k_\text{on}$}. For cell edges at the wound border, attachment occurs via protrusion into the nearest substrate vertex.
	
	\item Refine cell edges by subdividing edges longer than a maximum length, and merging edges shorter than a minimum length. This ensures an even distribution of adhesions, and allows the cells to assume curved shapes.
	
	\item Perform neighbour exchanges, also known as T1 transitions, when a cell edge shrinks below the threshold length, $L^*$, such that it lowers the total mechanical energy. {\color{black}Once an edge goes below the threshold length $L^*$, then that edge is replaced by a perpendicular contact of the same length.}
	
	\item Update modes of cell movement. Cells at the wound edge switch from crawling to purse string modes at a rate $k_\text{p}$. In the purse-string mode, cells can no longer crawl but instead carry a higher line tension around their wounded edge, $\gamma_\text{ps}$, modelling contractility of the actomyosin cable. Once cells are in the purse-string mode they remain so until wound closure or when the cell edge length shrinks to zero.	
	
	\item Move the cell vertices according to the overdamped equation of motion (Eq.~\eqref{eq:cell} or Eq.~\eqref{eq:cell2}). {\color{black}Individual nodes of the substrate spring mesh move at a velocity proportional to the net force resulting from focal adhesions and the gradient of mechanical energy of the spring mesh.}
\end{itemize}

\subsection*{Model parameters}
Table~\ref{table1} lists the parameters used in our simulations. {\color{black}The number of cells was chosen to be large enough to avoid finite size effects and displacement on the outer row of the cells. To confirm this, we ran wound healing simulations using different numbers of cells. As the number of cells increases from 50, closure time increases and then quickly plateaus after cell count reaches 100 (\nameref{S14_Fig}:A). We use a default value of 150 cells, but increase the cell number (in the range 150-250) while running simulations for wounds with larger sizes (Fig~\ref{fig:2}A). Substrate node density was chosen to be small enough so that a cell vertex is always close to a node in the substrate spring mesh, allowing focal adhesions to form with a relatively short length. As shown in \nameref{S14_Fig}:B, we find little dependence of closure time on node density, and use 0.6 $\mu$m$^{-2 }$ as the default value.}

The preferred area of the cell, $A_0$, is chosen to be approximately the same as the average area of MDCK cells in wound healing assays \cite{brugues2014, border_forces}. The preferred perimeter $P_0$ is chosen so that the cell shape index, $p_0 = P_0/\sqrt{A_0}$ is close to the {\color{black}value for a regular hexagon}, enabling us to study the effects {\color{black}of} cell shape anisotropy on wound healing speed. The substrate stiffness was chosen as a typical value for gels used in \textit{in vivo} wound healing assays \cite{brugues2014}; the Poisson's ratio of $1/3$ for the substrate is a consequence of using a triangular mesh of linear springs. The Young's modulus of the substrate defines the force scale in the simulations. {\color{black}The wound radius} was chosen {\color{black}to be in the range 5-30 $\mu$m}, similar to those in experimental studies \cite{brugues2014, border_forces, purse_string_closure}. 

Purse-string tension was estimated by taking the product of the force generated by a single myosin motor, $3$ pN \cite{miyata1995mechanical}, with the typical number of myosin motors in a contractile ring of length $15$ $\mu$m and thickness $1$ $\mu$m, $10^5$ \cite{biron2005molecular}, which gives a tension of $300$ nN. Next, we fit parameters for cell area and perimeter elasticities, $K$ and $\Gamma$, adhesion binding and unbinding rates, $k_\text{off}$ and $k_\text{on}$. Together, these parameters determine the overall tissue motility and the magnitude of traction force generation. Thus we fit them simulataneously to the experimental data for typical closure speed and traction force magnitudes generated during closure \cite{brugues2014, border_forces, purse_string_closure}. In addition, we examine the spatiotemporal pattern of traction forces generated during closure. For example, traction stresses are normally localized around the wound but are not evenly distributed around the perimeter. Low adhesion time leads to smooth closure but little traction force while higher adhesion binding times lead to an even distribution of traction around the wound but the closure dynamics are less smooth. 

We estimate the protrusion force, $f_p$, by comparing to single cell crawling speeds of $~15$ $\mu$m hr$^{-1}$ \cite{crawling_closure}. {\color{black}To this end, we simulated a single crawling cell with a fixed polarity vector, and calibrated $f_p$ to the value that resulted in a crawl speed of $~15$ $\mu$m hr$^{-1}$}. Internal motility speed was set to a similar value as cell crawling speeds. Dependence of wound closure time for variations in $f_p$ and $\gamma_\text{ps}$ are shown in {\color{black}\nameref{S2_Fig}:F}.  Whereas, the dependence of closure time on internal motility, $v_0$ is shown in {\color{black}\nameref{S8_Fig}:A,C}. The range of purse-string assembly rates were chosen so that the minimum value, $k_p = 0$ yields pure crawling, the maximum, $k_p = 1000$ hr$^{-1}$, yields 100\% purse-string coverage, and intermediate values produce a combination of purse-string and crawling.

\begin{table}[ht]
\centering
\caption{
{\bf Default Parameter Values}}
\begin{tabular}{|l+l|l|l|}
\hline
{\bf Parameter} & {\bf Default Value}\\ \thickhline
 \hline
 {\bf Cell} &\\
 \hline
Area elastic modulus, $K$  &  $0.2$ nN $\mu$m$^{-3}$\\ 

Preferred area, $A_0$ & $100\ \mu$m$^2$ \\
Preferred perimeter, $P_0$ & $36$ $\mu$m\\

Contractile tension, $\Gamma$ &  $20$ nN $\mu$m$^{-1}$\\

T1 threshold edge length, $L^*$ &  $1$ $\mu$m \\ 
Protrusion force, $f_p$ &  $2$ $\mu$N\\
Internal motility, $v_0$ &  $10$ $\mu$m hr$^{-1}$\\ 
Rotational diffusion, $D_r$ &  $5$ hr$^{-1}$\\ 
 \hline
 {\bf Substrate} &\\
 \hline
{\color{black}Node density in the spring mesh} & {\color{black}0.6 $\mu$m$^{-2}$}\\
Young's modulus, $E_s$ & $4$ kPa \\ 
Poisson's ratio, $\nu$ & 1/3 \\ 
Friction, $\mu$ &  $7.2 \times 10^4$ nN $\mu$m$^{-1}$s$^{-1}$ \\ 

Thickness, $h_s$ & $5$ $\mu$m \\ 

 Adhesion stiffness, $k_f$ & $4\times10^5$ nN $\mu$m$^{-1}$  \\

 Adhesion unbinding rate, $k_\text{off}$ &  $25$ hr$^{-1}$\\ 

  Adhesion binding rate, $k_\text{on}$ &  $500$ hr$^{-1}$\\ 

   \hline
 {\bf Wound} &\\
 \hline

  Radius & $15$ $\mu$m \\ 

  Aspect ratio & 1 \\ 
  Purse-string line tension, $\gamma_\text{ps}$ &  $300$ nN\\ 
 Purse-string transition rate, $k_\text{p}$ & $4$ hr$^{-1}$ \\ 
   \hline
 {\bf Other} &\\
 \hline

  Simulation timestep &  $3.6$ s\\ 
  Cell count & 150\\ \thickhline

\end{tabular}
\label{table1}
\end{table}

\subsection*{Traction stress computation}
We record displacements of the substrate mesh,  ${\bf u}={\bf r}-{\bf r}_0$, at each timestep during the simulation. These vectors are then interpolated to a square grid, from which strain is evaluated using the finite difference discretization of: $\epsilon_{kl} = \frac{1}{2}(\partial_k u_l + \partial_l u_k)$, where $k$ and $l$ are in-plane spatial coordinates.
The resultant stress is:{\color{black}
\begin{equation}\label{eq:stress}
\sigma_{kl} = \frac{E_s\nu}{(1+\nu)(1-2\nu)}\delta_{kl} \epsilon_{mm} + \frac{E_s}{(1+\nu)}\epsilon_{kl}\;.
\end{equation}}
The traction stress is calculated using $T_k = h_s\partial_l \sigma_{kl}$. {\color{black}The computed traction force vectors in the square grid are in excellent agreement with forces directly inferred from spring displacements in the triangular mesh (\nameref{S15_Fig})}. The strain energy density is given by $U = \frac{1}{2}\epsilon_{kl}\sigma_{kl}$.
For each simulation we calculate the mean strain energy as total strain energy transmitted to the substrate averaged over simulation time, $T$:
\begin{equation}\label{eq:strain_energy}
\langle \text{SE}\rangle = \frac{1}{T}\int_0^T dt\int_AdA\ h_s U(x,y,t)\;.
\end{equation}

{\color{black}
\subsection*{In vitro wound healing experiments}
Madin-Darby Canine Kidney (MDCK.2) cells (CRL-2936$^{\text{TM}}$; ATCC, Manassas, VA) were cultured in Eagle's Minimum Essential Medium (ATCC) containing 10\% fetal bovine serum (GIBCO Life Technologies) and 1\% penicillin/streptomycin at 37$^0$C and 5\% CO$_2$ in a  humidified incubator. MDCK.2 cells are stably transfected with a plasmid construct encoding for FTRActinEGFP (a gift from Sergey Plotnikov, University of Toronto).

Polyacrylamide gels are polymerized onto a glass coverslip at a ratio of 12\%:0.086\% polyacrylamide:bis-acrylamide to create a gel with an elastic modulus of 12.2 kPa~\cite{yeung2005}. After polymerization is complete, the polyacrylamide gels are reacted with 2mg/mL Sulfo-SANPAH (Thermo Fisher Scientific) and incubated with 1mg/mL Type 1 rat tail collagen (Corning, high concentration) for 2 hours in the dark~\cite{sabass2008}. Excess collagen is removed by rinsing with 1X Phosphate-buffered saline.

Confluent cell monolayers were grown on a polyacrylamide gel substrate with an elastic modulus of 12.2 kPa. Wounds were formed by laser ablation of a single cell using a 435 nm wavelength laser (Andor Technology, Belfast, Northern Ireland). Cell death causes monolayer retraction for $\sim$20 min after which the wounds close.
}


\section*{Supporting information}
\includegraphics[width=\columnwidth]{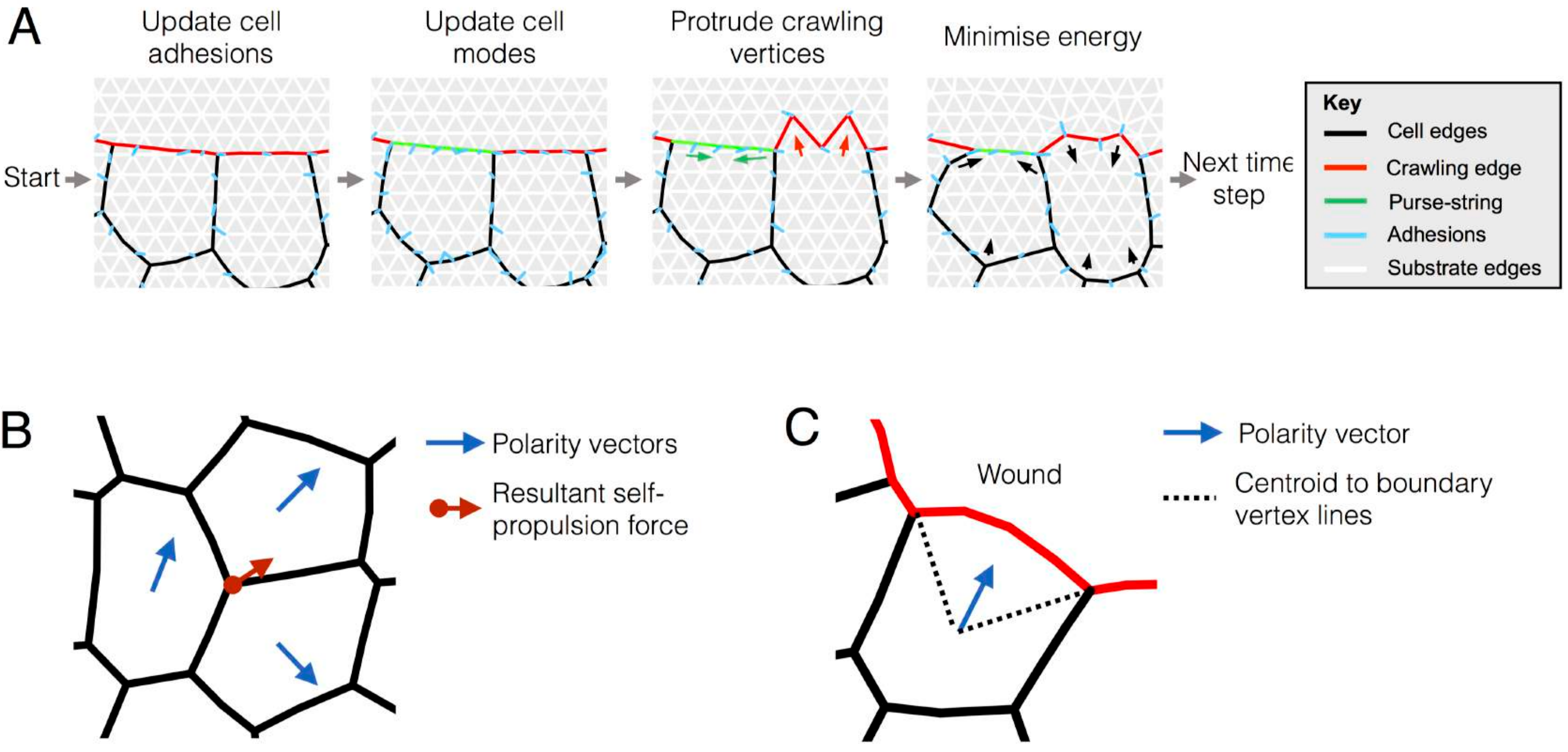}
\paragraph*{S1 Fig}
\label{S1_Fig}
{\bf Computational pipeline in the wound healing assay.} A: From left to right: 1) Update adhesion states for cell vertices. Adherent vertices attempt to unbind with a rate $k_\text{off}$, and unbound vertices attempt to bind to the nearest substrate mesh with a rate $k_\text{on}$. 2) Update cell modes from crawling (red) to purse-string (green) with a probability $k_\text{p}\Delta t$. 3) Protrude cell edges in crawling mode (red arrows) and contract cell edges on purse-string mode (green arrows). 4) Minimize mechanical energy to move the cell vertices down their mechanical energy gradient (black arrows). B: Illustration of self-propulsion force on a vertex in the bulk. The central vertex has a resultant force (red arrow) equal to the average force from its adjacent cells (blue arrows). C: The polarity vector (blue arrow) for a cell around the wounds bisects the angle between the lines from the cell centroid to the boundary vertices (dashed lines).

\includegraphics[width=\columnwidth]{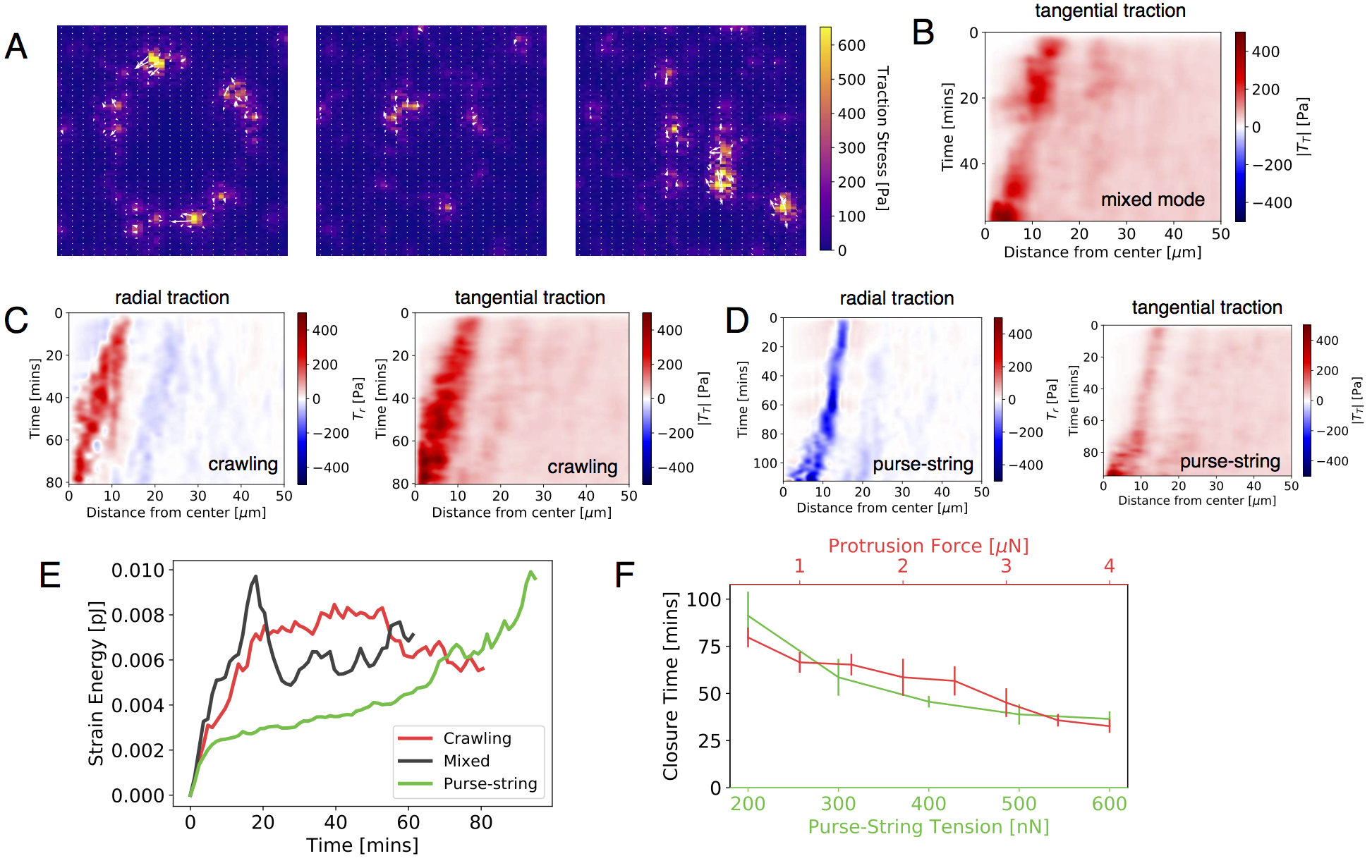}
\paragraph*{S2 Fig}
\label{S2_Fig}
{\bf Forces driving wound closure.} A: Traction stress distribution around a closing wound with $k_\text{p}=4$ hr$^{-1}$, at $t=5$ min (left), $t=30$ min (middle), $t=60$ min (right). {\color{black}B: Kymograph of tangential traction stress for the mixed mode of closure ($k_\text{p}=4$ hr$^{-1}$). C: Kymographs of radial  and tangential traction stress for the crawling ($k_\text{p}=0$ hr$^{-1}$) mode of closure. D: Kymographs of radial  and tangential traction stress for the purse-string ($k_\text{p}=1000$ hr$^{-1}$) mode of closure.} E: Total strain energy transmitted vs time for crawling, purse-string, and mixed modes of closure. F: Closure time as a function of purse-string tension (green) and protrusion force (red) for a mixed mode of closure ($k_\text{p}=4$ hr$^{-1}$). Error bars represent standard error of mean.

{\color{black}
\includegraphics[width=\columnwidth]{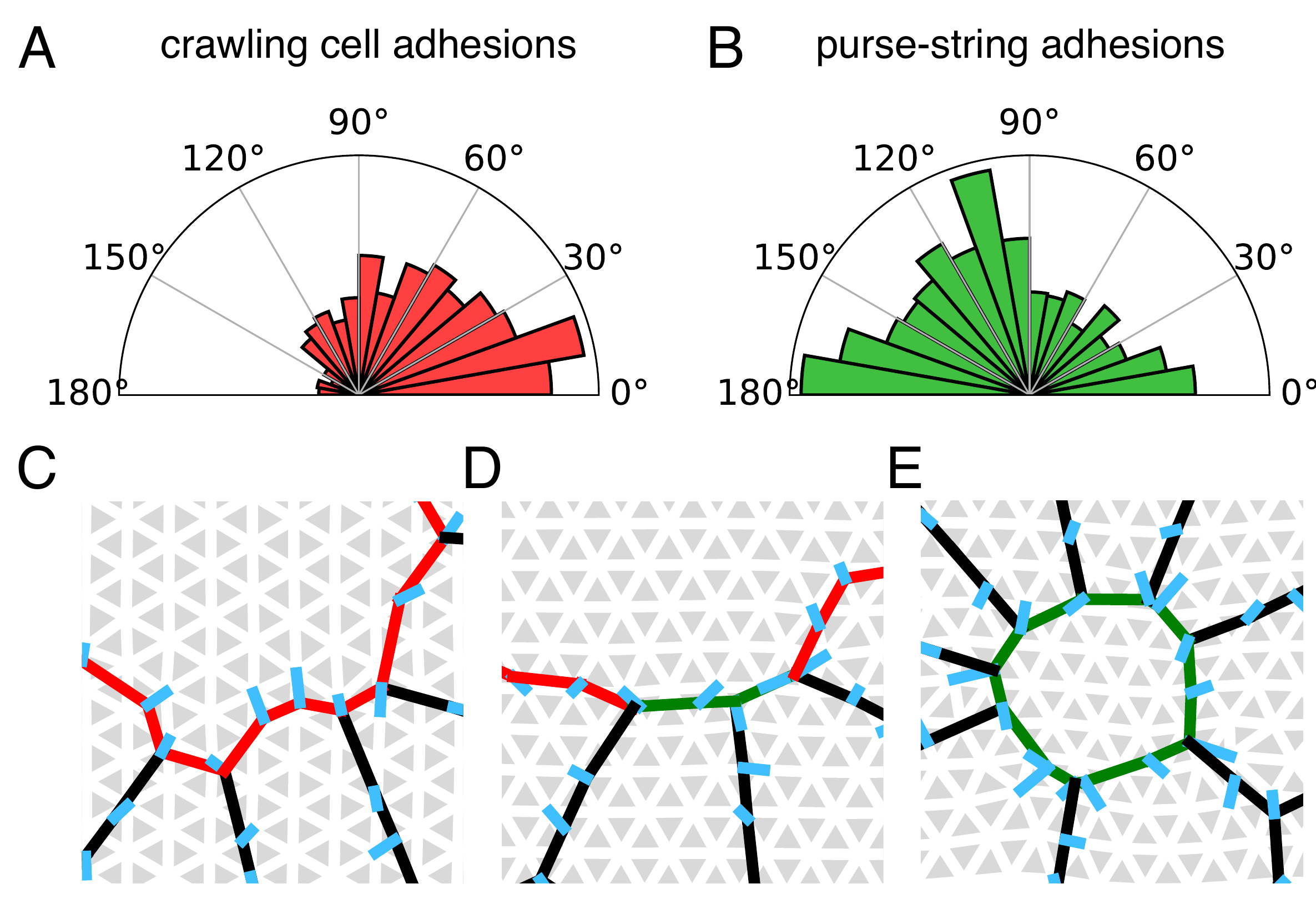}
\paragraph*{S3 Fig}
\label{S3_Fig}
{\bf Orientation of cell-substrate adhesions in leading edge cells.} Histograms of the local angle between cell-substrate bonds and the radial vector to the wound center, in (A) crawling and (B) purse-string cells at the leading edge. (C) Representative image of crawling cells with focal adhesions oriented normal to the wound edge. (D) A purse-string edge flanked between two crawling edges have its focal adhesions parallel to the wound edge. (E) Purse-string only wounds have a majority of adhesions oriented normal to the wound edge, due to normal driving forces arising from contractile tension in the purse-string. Green segments represent purse-string edges, while red segments are crawling cells.

\includegraphics[width=\columnwidth]{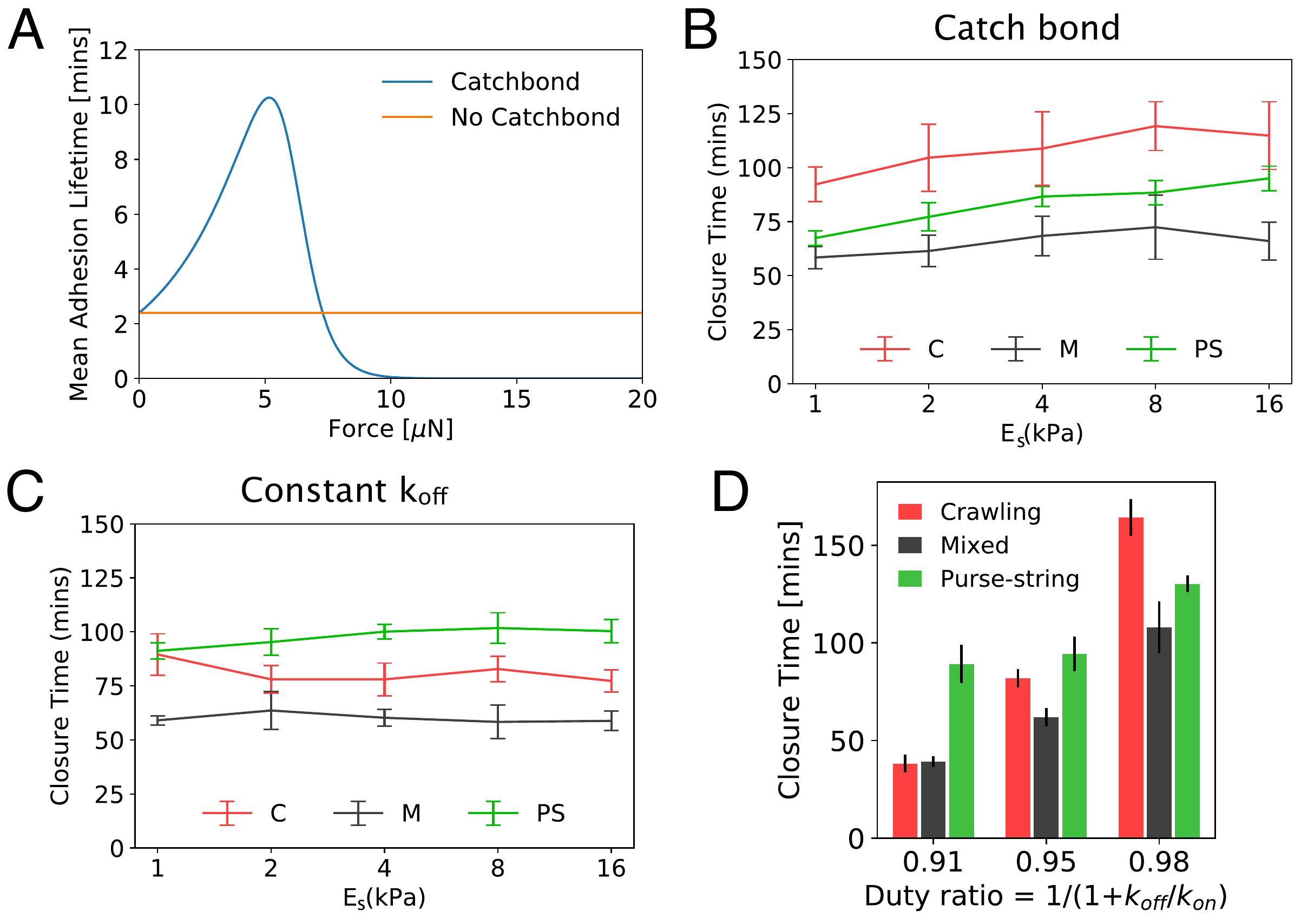}
\paragraph*{S4 Fig}
\label{S4_Fig}
{\bf Effect of cell-substrate adhesion kinetics on wound closure time.} (A) Mean adhesion lifetime, $k_\text{off}^{-1}$, vs applied force for a catch-bond model (blue) and constant $k_\text{off}$ (yellow). (B) Substrate stiffness dependence of wound closure time for a catch-bond model of cell-substrate adhesions, for crawling (red), purse-string (green) and mixed ($k_p=4$ hr$^{-1}$, black) modes of closure. (C) Wound closure time vs substrate stiffness for constant $k_\text{off}$. Each data point represent average over 6 simulations. Error bars show standard deviation. (D) Closure time vs duty ratio of focal adhesion bonds, $k_\text{on}/(k_\text{off}+k_\text{on})$, for crawling, purse-string and mixed modes of wound closure. Duty ratio is varied by changing the detachment rate, $k_\text{off}$, for a fixed $k_\text{on}$.

}
\includegraphics[width=\columnwidth]{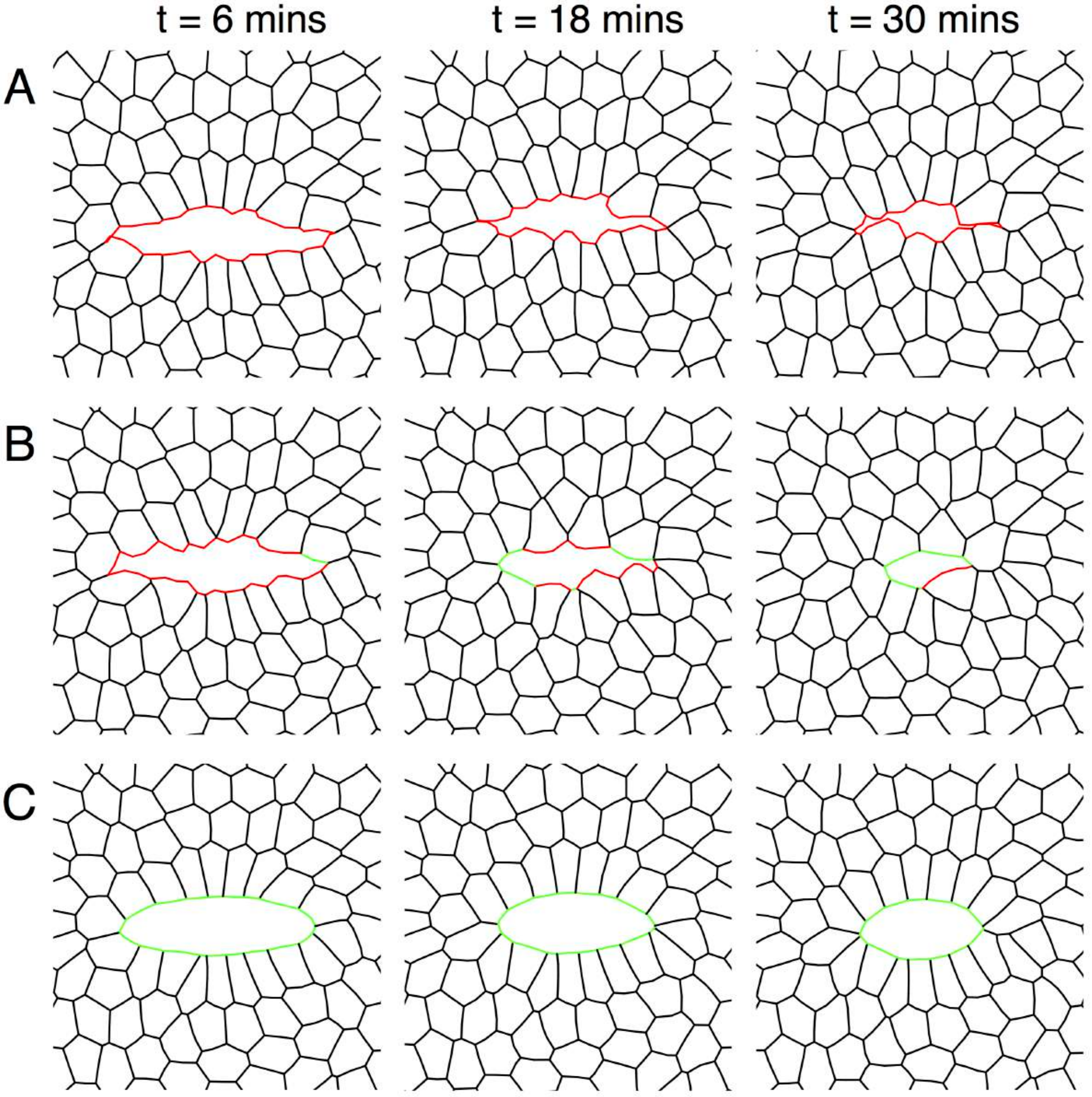}
\paragraph*{S5 Fig}
\label{S5_Fig}
{\bf Shape dependent dynamics of wound closure.} Wound morphologies for (A) crawling ($k_\text{p}=0$ hr$^{-1}$), (B) mixed ($k_\text{p}=4$ hr$^{-1}$), and (C) purse-string ($k_\text{p}=1000$ hr$^{-1}$) modes of closure, at $t=6$ min (left), $t=18$ min (middle), $t=30$ min (right). The initial aspect ratio of the wound is $4$. 

\includegraphics[width=\columnwidth]{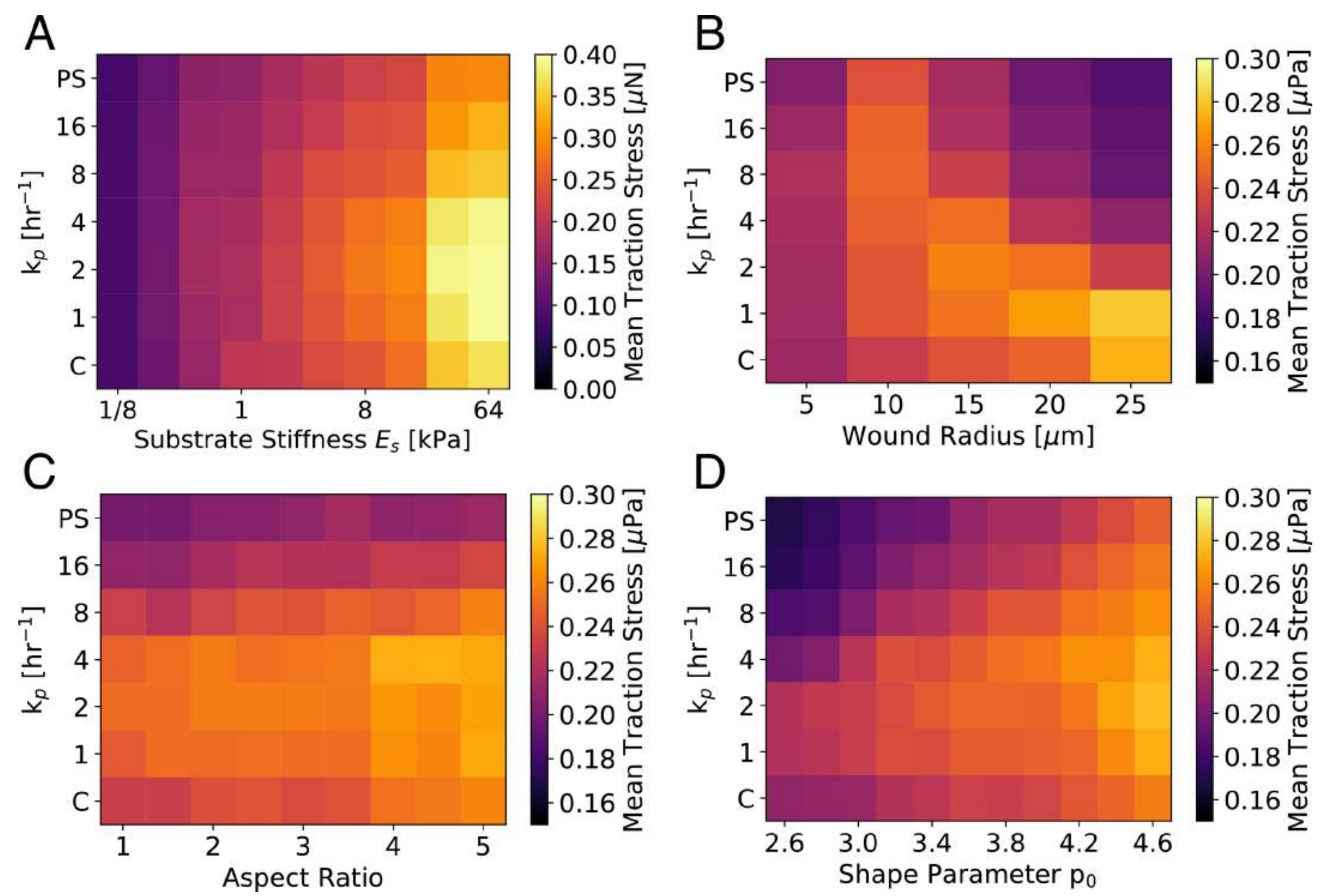}
\paragraph*{S6 Fig}
\label{S6_Fig}
{\bf Dependence of traction stress on cell, substrate and wound properties.} Temporal mean of spatially averaged traction stress during wound closure for different values of $k_{p}$ and (A) substrate stiffness, (B) wound radius, (C) wound aspect ratio, and (D) shape parameter $p_0$.

{\color{black}
\includegraphics[width=\columnwidth]{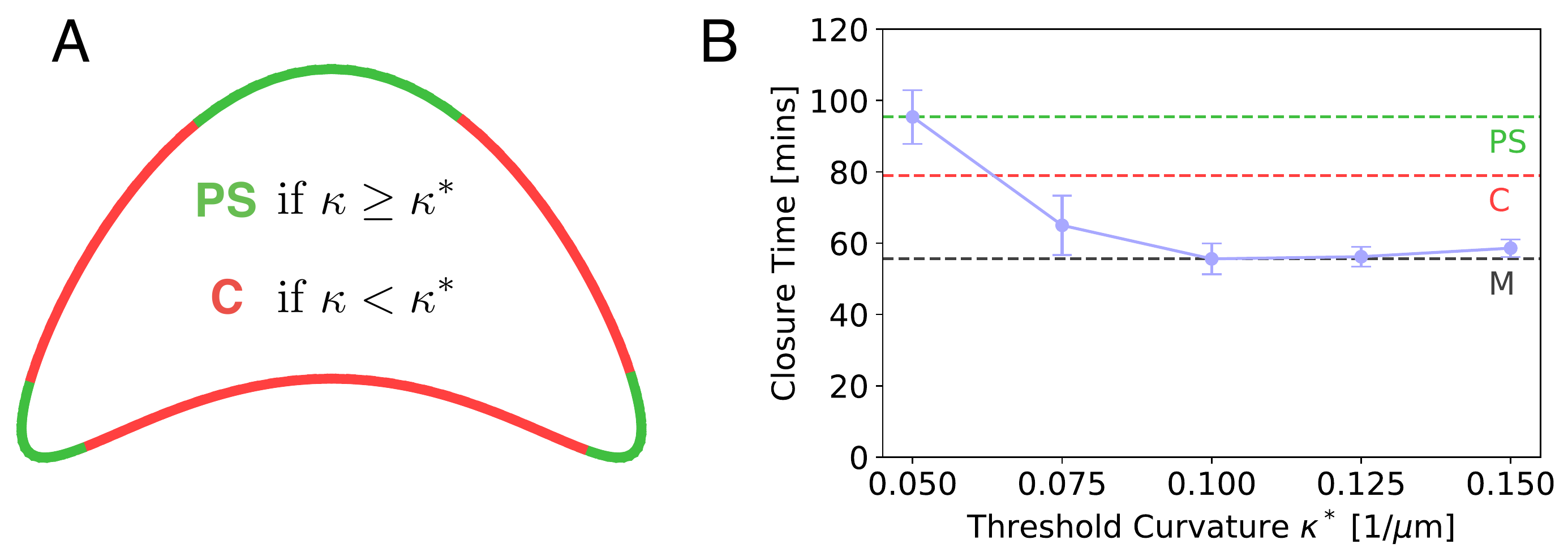}
\paragraph*{S7 Fig}
\label{S7_Fig}
{\bf Model for curvature dependent purse-string formation.} (A) Schematic showing purse-string and crawling edges for a wound with non-uniform curvature. Purse-string (PS; green) forms on leading edges with curvature $\kappa>\kappa^*$, where $\kappa^*$ is a threshold curvature. Cells prefer to crawl (C; red) if $\kappa<\kappa^*$. (B) Wound closure time vs $\kappa^*$ for the concave shaped wound in (A). The optimum threshold curvature is chosen to be the one that minimizes wound closure time. Dashed lines indicate wound closure times for pure crawling (red), pure purse-string (red) and stochastic mixed (black) modes of closure. 
}

\includegraphics[width=\columnwidth]{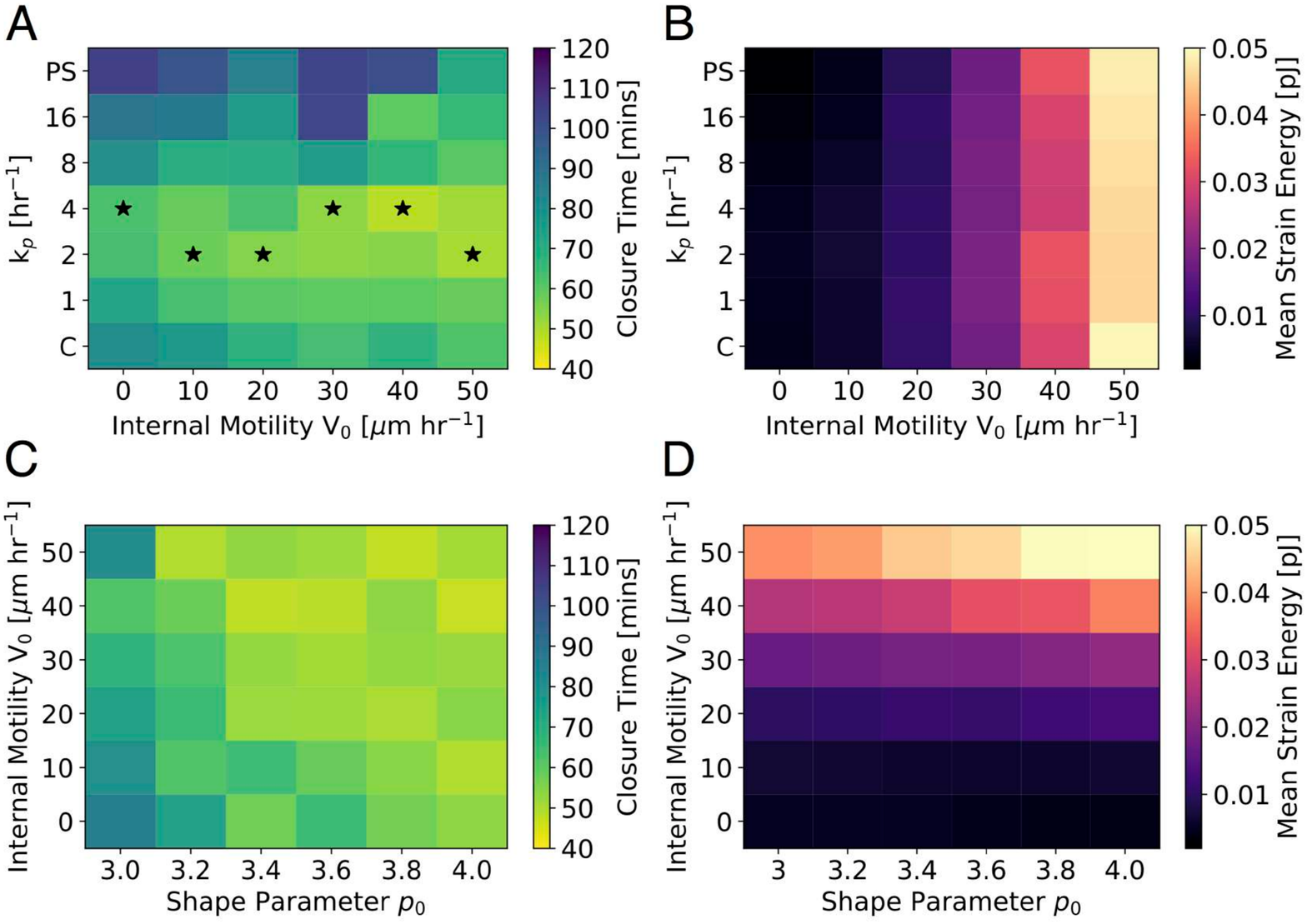}
\paragraph*{S8 Fig}
\label{S8_Fig}
{\bf Internal motility accelerates the rate of wound closure.} (A) Closure time, and (B) mean strain energy for different values of internal motility $v_0$ and $k_{p}$. Starred cells indicate the fastest wound closure for a given $v_0$ with varying purse-string assembly rates. (C) Closure time, and (D) average strain energy for different values of shape parameter $p_0$ and internal motility $v_0$, for a mixed mode of closure ($k_\text{p}=4$ hr$^{-1}$).

\includegraphics[width=\columnwidth]{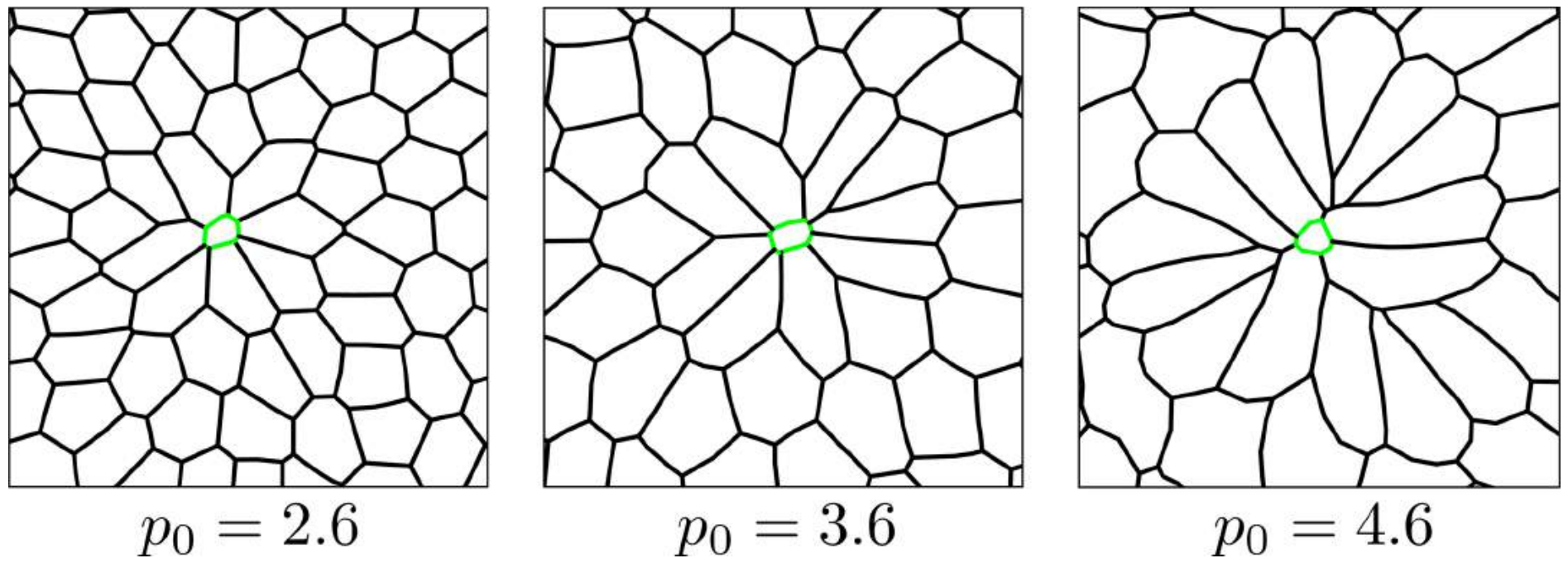}
\paragraph*{S9 Fig}
\label{S9_Fig}
{\bf Tissue morphology prior to wound closure} for different values of shape parameter $p_0$, from 3.0 (solid-like tisue) to 4.0 (fluid-like tissue).

{\color{black}
\includegraphics[width=\columnwidth]{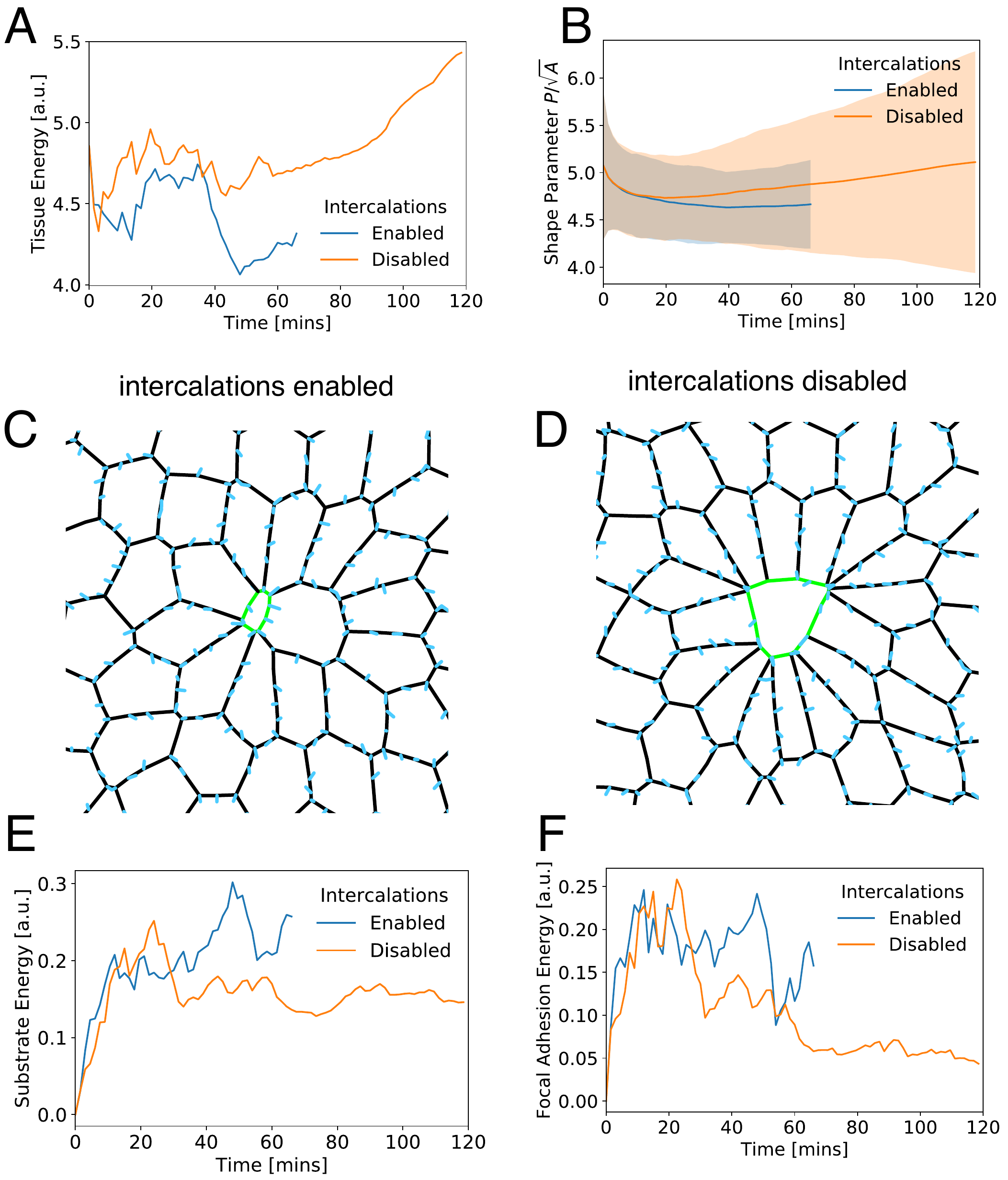}
\paragraph*{S10 Fig}
\label{S10_Fig}
{\bf Intercalations reduce tissue mechanical energy.} (A) Total tissue mechanical energy vs time, with intercalations enabled and disabled during wound closure. (B) Mean cell shape parameter vs time. Shaded regions represent one standard deviation. With intercalations disabled, cells elongate and have more variability in shape. (C-D) Simulation image showing tissue morphology before closure with intercalations (C), and in a jammed state without intercalations (D). Cells are much more elongated when intercalations are disabled. (E) Total substrate strain energy, and (F) total focal adhesion strain energy, over time with intercalations enabled and disabled.
}

\includegraphics[width=\columnwidth]{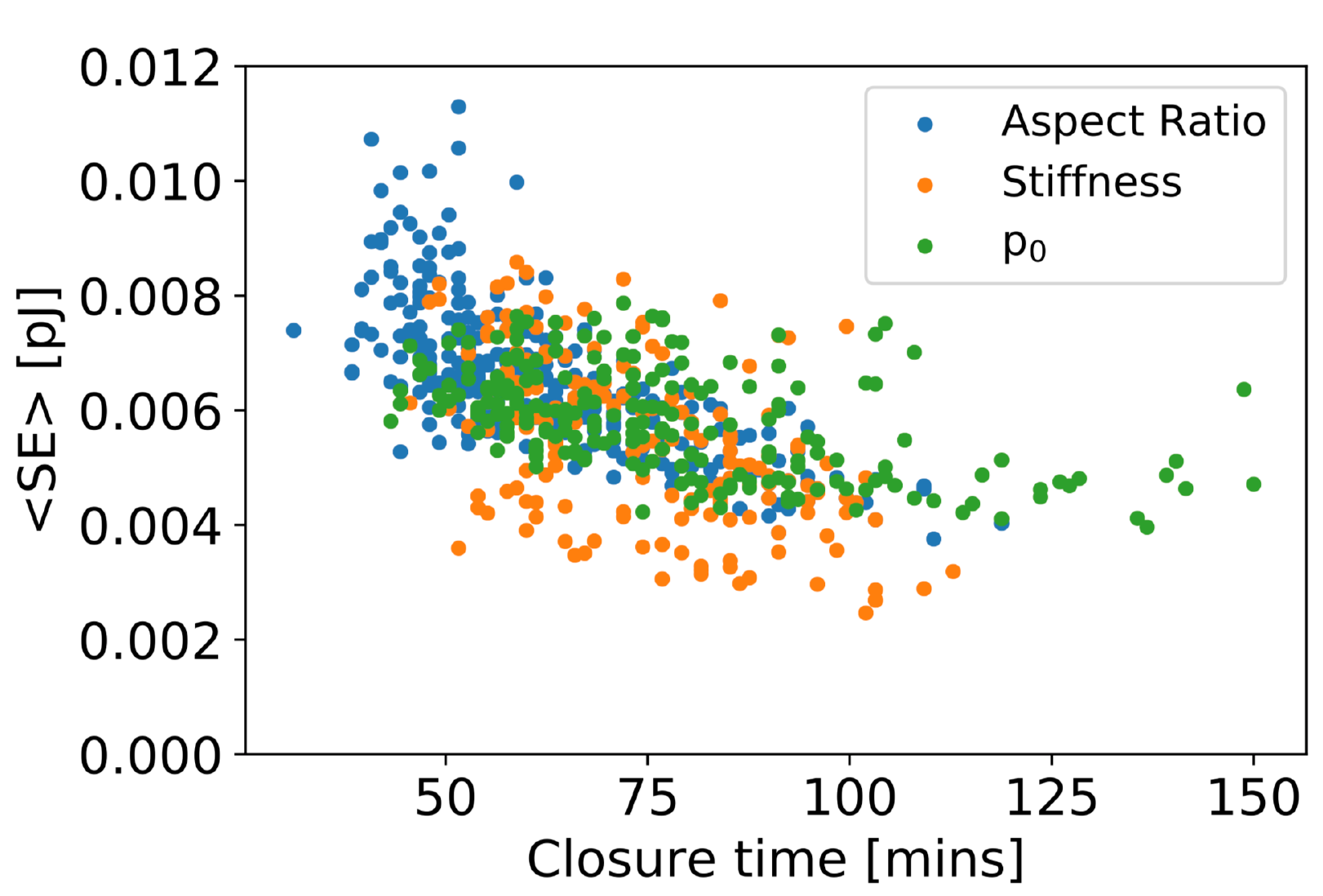}
\paragraph*{S11 Fig}
\label{S11_Fig}
{\bf Faster closure leads to higher strain energy transmitted to the substrate.} Mean strain energy vs closure time. Each data point represents a different simulation. The color corresponds to the parameter that was being varied in that simulation.
{\color{black}
\includegraphics[width=\columnwidth]{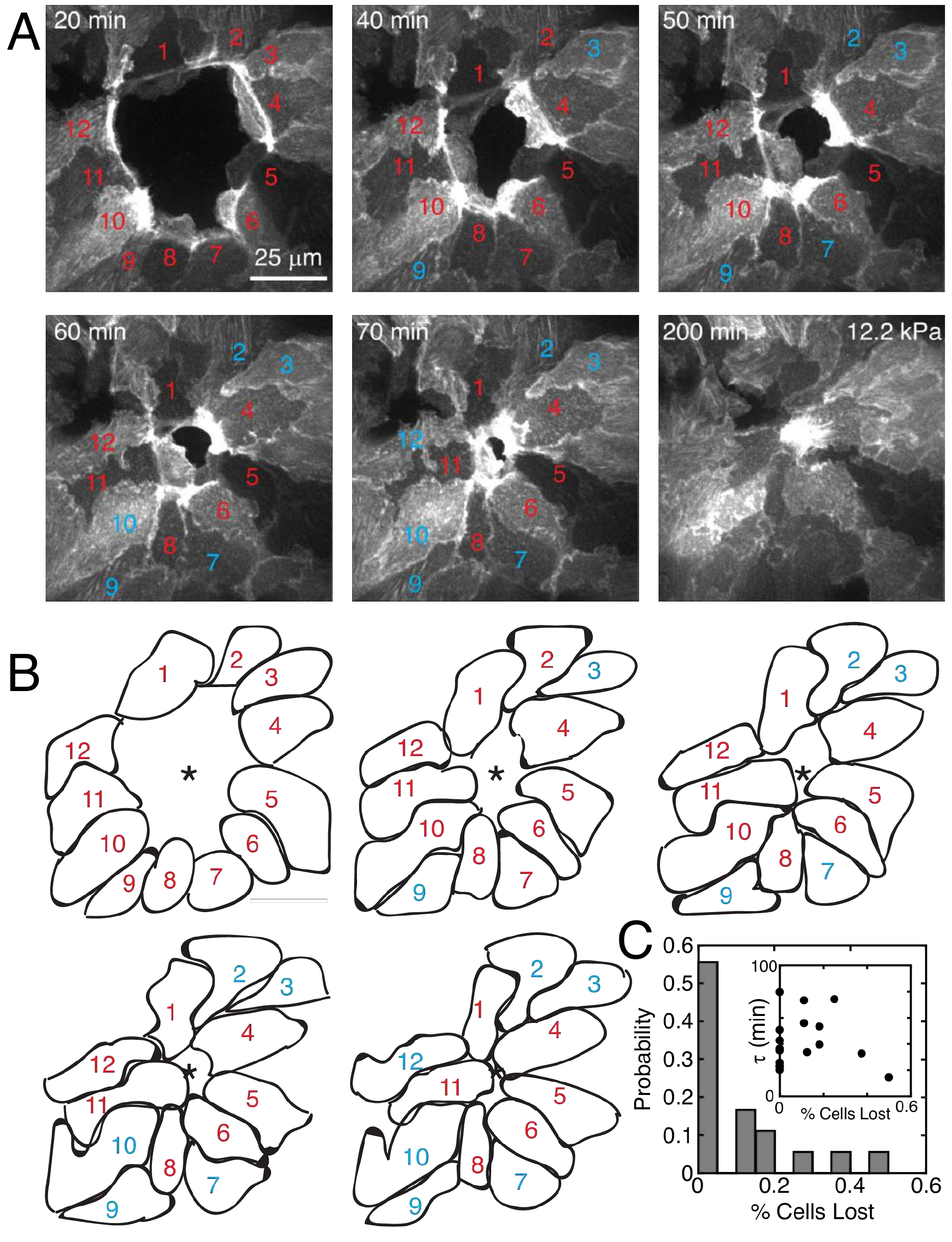}
\paragraph*{S12 Fig}
\label{S12_Fig}
{\bf Intercalation events can occur during wound healing}. (A) Time-lapse images of fluorescent F-actin within MDCK cells closing a wound and (B) the drawn outlines of cells initially at the leading edge. Cells at the leading edge at each time point are numbered in red, whereas cells excluded from the leading edge during closure are numbered in cyan. (C) The probability distribution of fractional cell loss for $N=18$ wounds where the average number of cells initially at the leading edge is $9\pm 2$. $N=8$ wounds exhibit a loss of cells at the leading edge during closure. Within this subset, the average percentage of cells lost is $0.23 \pm 0.14$. (C-inset) The closure timescale, $\tau$, calculated from $A(t)=A(0)e^{-t/\tau}$, where $A(t)$ is the area of the wound at time $t$, vs the fractional cell loss at the leading edge.

\includegraphics[width=\columnwidth]{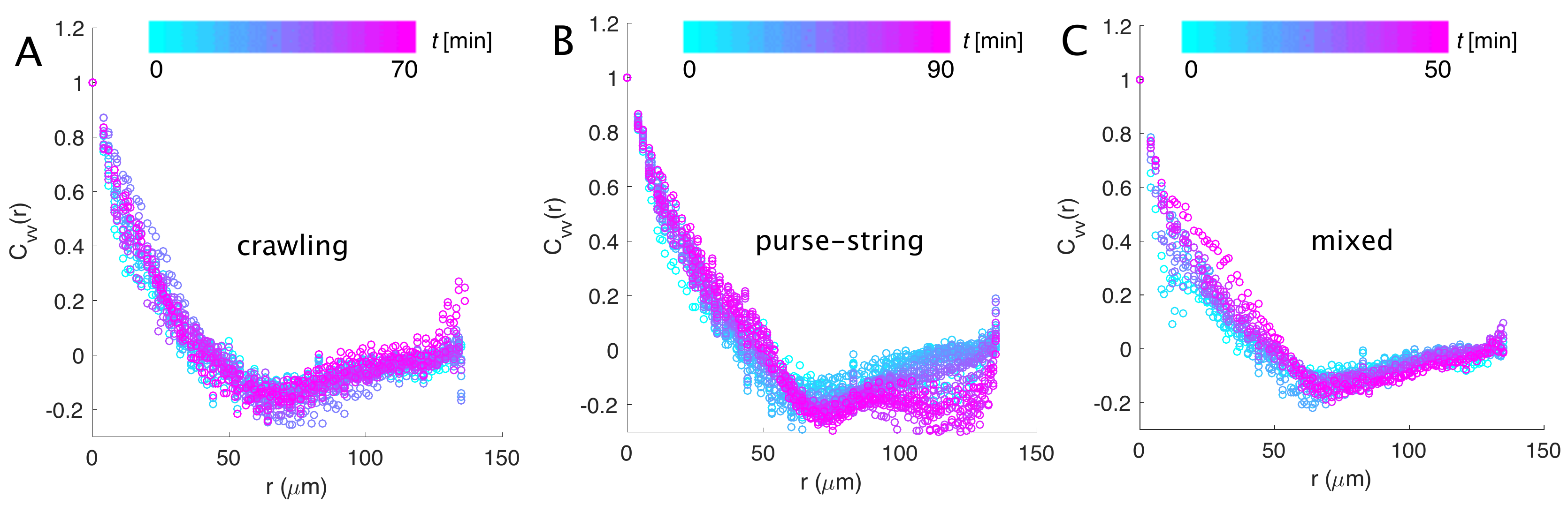}
\paragraph*{S13 Fig}
\label{S13_Fig}
{\bf Velocity-velocity correlation of cells during wound closure.} Figure shows velocity-velocity correlation function, $C_{vv}(r)=\langle {\bf v}(0).{\bf v}(r)\rangle/\langle {\bf v}(0)^2\rangle$, where $r$ is the distance between two cell center velocity vectors, ${\bf v}$. $C_{vv}(r)$ is shown at different time points (indicated by color) for (A) crawling, (B) purse-string, and (C) mixed modes of wound closure. Velocity vectors of cells on opposite sides of the wound are anti-correlated.
  
\includegraphics[width=\columnwidth]{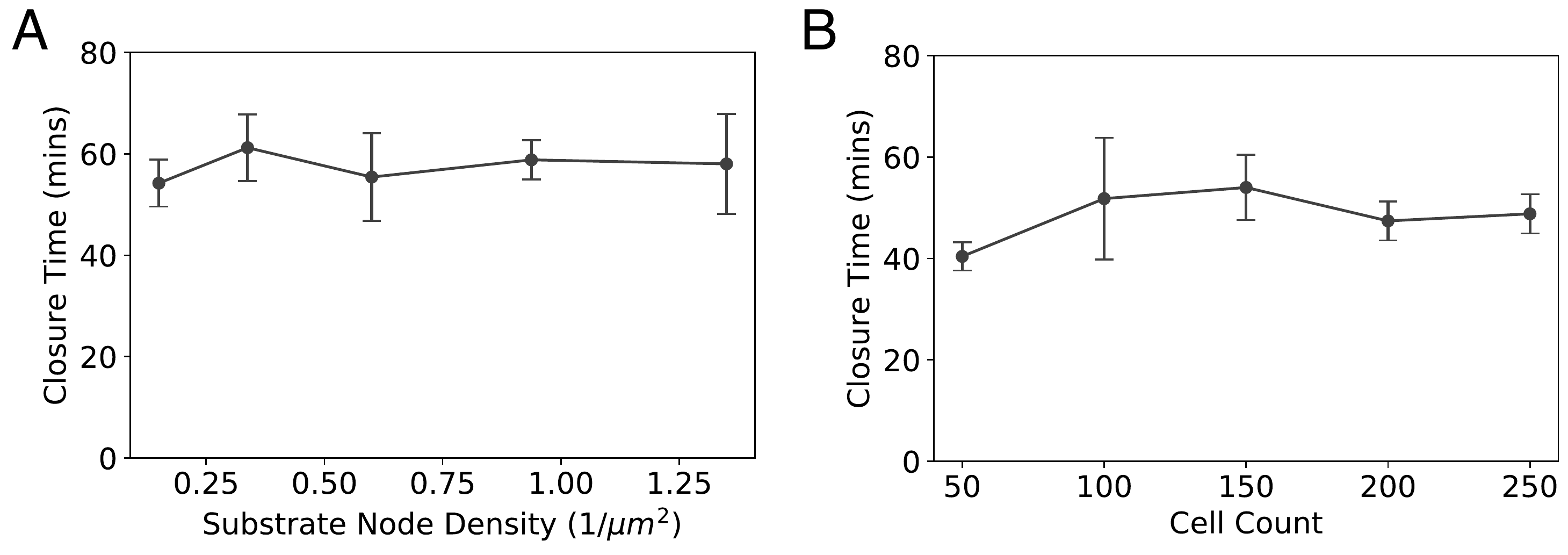}
\paragraph*{S14 Fig}
\label{S14_Fig}
{\bf Dependence of wound closure time on cell count and substrate node density.} Closure time vs (A) cell count, (B) density of nodes in the substrate spring mesh, for a wound of fixed initial size.

\includegraphics[width=\columnwidth]{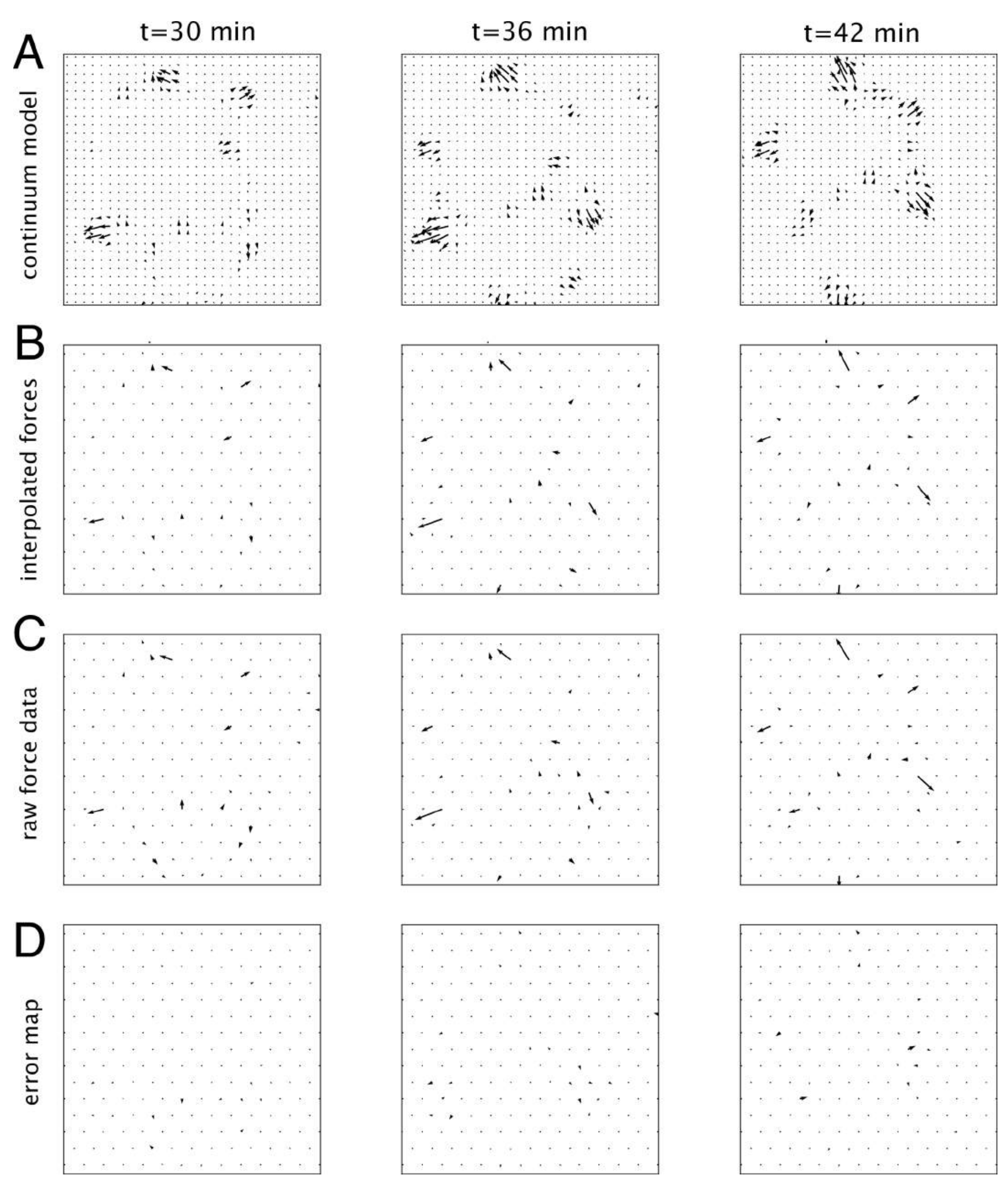}
\paragraph*{S15 Fig}
\label{S15_Fig}
{\bf Comparison of traction force computation methods.} Figure shows traction force vectors using two different methods computed during wound closure at $t=30$ min (left column), $t=36$ min (middle column), and $t=42$ min (right column). (A) Traction force vectors computed using the continuum elasticity equation (9). (B) Continuum model based forces in (A) interpolated on the substrate triangular mesh. (C) Traction forces directly computed from displacements in the substrate spring mesh. (D) Error map showing the difference of traction force vectors in (B) and (C). Lengths of arrows are proportional to the magnitude of the traction force, and the scale is consistent between images.
}

\paragraph*{S1 Video.}
\label{S1_Video}
{\bf Wound healing driven by a mixture of crawling and purse-string.}  

\paragraph*{S2 Video.}
\label{S2_Video}
{\bf Wound healing driven by pure purse-string}

\paragraph*{S3 Video.}
\label{S3_Video}
{\bf Wound healing driven by pure cell crawling.} 

\paragraph*{S4 Video.}
\label{S4_Video}
{\bf Wound closure simulations for a circular and an elliptical wound.}

\paragraph*{S5 Video.}
\label{S5_Video}
{\bf Wound closure simulations for a concave wound shape.}

\paragraph*{S6 Video.}
\label{S6_Video}
{\bf Effect of tissue fluidity on wound closure.} Left: $p_0=2.6$, Right: $p_0=4.6$.


\begin{thebibliography}{10}

\bibitem{friedl2009}
Friedl P, Gilmour D.
\newblock Collective cell migration in morphogenesis, regeneration and cancer.
\newblock Nature reviews Molecular cell biology. 2009;10(7):445.

\bibitem{grose2002}
Grose R, Hutter C, Bloch W, Thorey I, Watt FM, F{\"a}ssler R, et~al.
\newblock A crucial role of $\beta$1 integrins for keratinocyte migration in
  vitro and during cutaneous wound repair.
\newblock Development. 2002;129(9):2303--2315.

\bibitem{jacinto2002dynamic}
Jacinto A, Woolner S, Martin P.
\newblock Dynamic analysis of dorsal closure in Drosophila: from genetics to
  cell biology.
\newblock Developmental Cell. 2002;3(1):9--19.

\bibitem{rosenblatt2001epithelial}
Rosenblatt J, Raff MC, Cramer LP.
\newblock An epithelial cell destined for apoptosis signals its neighbors to
  extrude it by an actin-and myosin-dependent mechanism.
\newblock Current Biology. 2001;11(23):1847--1857.

\bibitem{begnaud2016}
Begnaud S, Chen T, Delacour D, M{\`e}ge RM, Ladoux B.
\newblock Mechanics of epithelial tissues during gap closure.
\newblock Current Opinion in Cell Biology. 2016;42:52--62.

\bibitem{martin1992actin}
Martin P, Lewis J.
\newblock Actin cables and epidermal movement in embryonic wound healing.
\newblock Nature. 1992;360(6400):179--183.

\bibitem{crawling_closure}
Anon E, Serra-Picamal X, Hersen P, Gauthier NC, Sheetz MP, Trepat X, et~al.
\newblock Cell crawling mediates collective cell migration to close undamaged
  epithelial gaps.
\newblock Proceedings of the National Academy of Sciences.
  2012;109(27):10891--10896.
\newblock doi:{10.1073/pnas.1117814109}.

\bibitem{fenteany2000}
Fenteany G, Janmey PA, Stossel TP.
\newblock Signaling pathways and cell mechanics involved in wound closure by
  epithelial cell sheets.
\newblock Current Biology. 2000;10(14):831--838.

\bibitem{bement1993}
Bement WM, Forscher P, Mooseker MS.
\newblock A novel cytoskeletal structure involved in purse string wound closure
  and cell polarity maintenance.
\newblock The Journal of Cell Biology. 1993;121(3):565--578.

\bibitem{trepat2009}
Trepat X, Wasserman MR, Angelini TE, Millet E, Weitz DA, Butler JP, et~al.
\newblock Physical forces during collective cell migration.
\newblock Nature physics. 2009;5(6):426.

\bibitem{tambe2011}
Tambe DT, Hardin CC, Angelini TE, Rajendran K, Park CY, Serra-Picamal X, et~al.
\newblock Collective cell guidance by cooperative intercellular forces.
\newblock Nature materials. 2011;10(6):469.

\bibitem{purse_string_closure}
Vedula SRK, Peyret G, Cheddadi I, Chen T, Brugu{\'e}s A, Hirata H, et~al.
\newblock Mechanics of epithelial closure over non-adherent environments.
\newblock Nature Communications. 2015;6:6111.

\bibitem{border_forces}
Cochet-Escartin O, Ranft J, Silberzan P, Marcq P.
\newblock Border forces and friction control epithelial closure dynamics.
\newblock Biophysical Journal. 2014;106(1):65--73.

\bibitem{ladoux2015geometry}
Ravasio A, Cheddadi I, Chen T, Pereira T, Ong HT, Bertocchi C, et~al.
\newblock Gap geometry dictates epithelial closure efficiency.
\newblock Nature Communications. 2015;6.

\bibitem{sherratt1992}
Sherratt JA, Martin P, Murray J, Lewis J.
\newblock Mathematical models of wound healing in embryonic and adult
  epidermis.
\newblock Mathematical Medicine and Biology: A Journal of the IMA.
  1992;9(3):177--196.

\bibitem{lee2011}
Lee P, Wolgemuth CW.
\newblock Crawling cells can close wounds without purse strings or signaling.
\newblock PLoS Computational Biology. 2011;7(3):e1002007.

\bibitem{banerjee2015}
Banerjee S, Utuje KJ, Marchetti MC.
\newblock Propagating stress waves during epithelial expansion.
\newblock Physical Review Letters. 2015;114(22):228101.

\bibitem{nagai2009computer}
Nagai T, Honda H.
\newblock Computer simulation of wound closure in epithelial tissues:
  Cell--basal-lamina adhesion.
\newblock Physical Review E. 2009;80(6):061903.

\bibitem{salm2012}
Salm M, Pismen L.
\newblock Chemical and mechanical signaling in epithelial spreading.
\newblock Physical biology. 2012;9(2):026009.

\bibitem{basan2013}
Basan M, Elgeti J, Hannezo E, Rappel WJ, Levine H.
\newblock Alignment of cellular motility forces with tissue flow as a mechanism
  for efficient wound healing.
\newblock Proceedings of the National Academy of Sciences.
  2013;110(7):2452--2459.

\bibitem{mirams2013}
Mirams GR, Arthurs CJ, Bernabeu MO, Bordas R, Cooper J, Corrias A, et~al.
\newblock Chaste: an open source C++ library for computational physiology and
  biology.
\newblock PLoS Computational Biology. 2013;9(3):e1002970.

\bibitem{banerjee2018}
Banerjee S, Marchetti MC.
\newblock Continuum models of collective cell migration.
\newblock arXiv preprint arXiv:180506531. 2018;.

\bibitem{kopf2013}
K{\"o}pf MH, Pismen LM.
\newblock A continuum model of epithelial spreading.
\newblock Soft Matter. 2013;9(14):3727--3734.

\bibitem{graner1992}
Graner F, Glazier JA.
\newblock Simulation of biological cell sorting using a two-dimensional
  extended Potts model.
\newblock Physical review letters. 1992;69(13):2013.

\bibitem{Albert2016}
Albert PJ, Schwarz US.
\newblock Dynamics of cell ensembles on adhesive micropatterns: bridging the
  gap between single cell spreading and collective cell migration.
\newblock PLoS Computational Biology. 2016;12(4):e1004863.

\bibitem{honda1980}
Honda H, Eguchi G.
\newblock How much does the cell boundary contract in a monolayered cell sheet?
\newblock Journal of Theoretical Biology. 1980;84(3):575--588.

\bibitem{farhadifar2007}
Farhadifar R, R{\"o}per JC, Aigouy B, Eaton S, J{\"u}licher F.
\newblock The influence of cell mechanics, cell-cell interactions, and
  proliferation on epithelial packing.
\newblock Current Biology. 2007;17(24):2095--2104.

\bibitem{lober2015}
L{\"o}ber J, Ziebert F, Aranson IS.
\newblock Collisions of deformable cells lead to collective migration.
\newblock Scientific reports. 2015;5:9172.

\bibitem{tarle2015}
Tarle V, Ravasio A, Hakim V, Gov NS.
\newblock Modeling the finger instability in an expanding cell monolayer.
\newblock Integrative Biology. 2015;7(10):1218--1227.

\bibitem{zimmermann2016}
Zimmermann J, Camley BA, Rappel WJ, Levine H.
\newblock Contact inhibition of locomotion determines cell--cell and
  cell--substrate forces in tissues.
\newblock Proceedings of the National Academy of Sciences.
  2016;113(10):2660--2665.

\bibitem{brugues2014}
Brugues A, Anon E, Conte V, Veldhuis JH, Gupta M, Colombelli J, et~al.
\newblock Forces driving epithelial wound healing.
\newblock Nature Physics. 2014;10(9):683--690.

\bibitem{nagai_vm}
Nagai T, Honda H.
\newblock Wound healing mechanism in epithelial tissues cell adhesion to basal
  lamina.
\newblock WSEAS Transactions on Biology and Biomedicine. 2006;3(6):389.

\bibitem{fletcher2014}
Fletcher AG, Osterfield M, Baker RE, Shvartsman SY.
\newblock Vertex models of epithelial morphogenesis.
\newblock Biophysical Journal. 2014;106(11):2291--2304.

\bibitem{manning2010}
Manning ML, Foty RA, Steinberg MS, Schoetz EM.
\newblock Coaction of intercellular adhesion and cortical tension specifies
  tissue surface tension.
\newblock Proceedings of the National Academy of Sciences.
  2010;107(28):12517--12522.

\bibitem{vm_shapes}
Staple D, Farhadifar R, R{\"o}per JC, Aigouy B, Eaton S, J{\"u}licher F.
\newblock Mechanics and remodelling of cell packings in epithelia.
\newblock The European Physical Journal E. 2010;33(2):117--127.

\bibitem{ajeti2018}
Ajeti V, Tabatabai AP, Fleszar AJ, Staddon MF, Seara DS, Suarez C, et~al.
\newblock Epithelial Wound Healing Coordinates Distinct Actin Network
  Architectures to Conserve Mechanical Work and Balance Power.
\newblock arXiv preprint arXiv:180606768. 2018;.

\bibitem{discher_stiffness}
Discher DE, Janmey P, Wang Yl.
\newblock Tissue cells feel and respond to the stiffness of their substrate.
\newblock Science. 2005;310(5751):1139--1143.

\bibitem{schwarz2013}
Schwarz US, Safran SA.
\newblock Physics of adherent cells.
\newblock Reviews of Modern Physics. 2013;85(3):1327.

\bibitem{walcott2010}
Walcott S, Sun SX.
\newblock A mechanical model of actin stress fiber formation and substrate
  elasticity sensing in adherent cells.
\newblock Proceedings of the National Academy of Sciences.
  2010;107(17):7757--7762.

\bibitem{kong2009}
Kong F, Garc{\'\i}a AJ, Mould AP, Humphries MJ, Zhu C.
\newblock Demonstration of catch bonds between an integrin and its ligand.
\newblock The Journal of Cell Biology. 2009;185(7):1275--1284.

\bibitem{pereverzev2005}
Pereverzev YV, Prezhdo OV, Forero M, Sokurenko EV, Thomas WE.
\newblock The two-pathway model for the catch-slip transition in biological
  adhesion.
\newblock Biophysical Journal. 2005;89(3):1446--1454.

\bibitem{foty2005}
Foty RA, Steinberg MS.
\newblock The differential adhesion hypothesis: a direct evaluation.
\newblock Developmental Biology. 2005;278(1):255--263.

\bibitem{mertz2012}
Mertz AF, Banerjee S, Che Y, German GK, Xu Y, Hyland C, et~al.
\newblock Scaling of traction forces with the size of cohesive cell colonies.
\newblock Physical Review Letters. 2012;108(19):198101.

\bibitem{mertz2013}
Mertz AF, Che Y, Banerjee S, Goldstein JM, Rosowski KA, Revilla SF, et~al.
\newblock Cadherin-based intercellular adhesions organize epithelial
  cell--matrix traction forces.
\newblock Proceedings of the National Academy of Sciences.
  2013;110(3):842--847.

\bibitem{bi2015}
Bi D, Lopez J, Schwarz J, Manning ML.
\newblock A density-independent rigidity transition in biological tissues.
\newblock Nature Physics. 2015;11(12):1074.

\bibitem{ranft2010}
Ranft J, Basan M, Elgeti J, Joanny JF, Prost J, J{\"u}licher F.
\newblock Fluidization of tissues by cell division and apoptosis.
\newblock Proceedings of the National Academy of Sciences.
  2010;107(49):20863--20868.

\bibitem{bi2016}
Bi D, Yang X, Marchetti MC, Manning ML.
\newblock Motility-driven glass and jamming transitions in biological tissues.
\newblock Physical Review X. 2016;6(2):021011.

\bibitem{barton2017}
Barton DL, Henkes S, Weijer CJ, Sknepnek R.
\newblock Active Vertex Model for cell-resolution description of epithelial
  tissue mechanics.
\newblock PLoS Computational Biology. 2017;13(6):e1005569.

\bibitem{razzell2014}
Razzell W, Wood W, Martin P.
\newblock Recapitulation of morphogenetic cell shape changes enables wound
  re-epithelialisation.
\newblock Development. 2014;141(9):1814--1820.

\bibitem{curran2017}
Curran S, Strandkvist C, Bathmann J, de~Gennes M, Kabla A, Salbreux G, et~al.
\newblock Myosin II controls junction fluctuations to guide epithelial tissue
  ordering.
\newblock Developmental cell. 2017;43(4):480--492.

\bibitem{wyatt2015}
Wyatt TP, Harris AR, Lam M, Cheng Q, Bellis J, Dimitracopoulos A, et~al.
\newblock Emergence of homeostatic epithelial packing and stress dissipation
  through divisions oriented along the long cell axis.
\newblock Proceedings of the National Academy of Sciences.
  2015;112(18):5726--5731.

\bibitem{szabo2010}
Szab{\'o} A, {\"U}nnep R, M{\'e}hes E, Twal W, Argraves W, Cao Y, et~al.
\newblock Collective cell motion in endothelial monolayers.
\newblock Physical Biology. 2010;7(4):046007.

\bibitem{schaumann2018}
Schaumann EN, Staddon MF, Gardel ML, Banerjee S.
\newblock Force localization modes in dynamic epithelial colonies.
\newblock bioRxiv. 2018; p. 336164.

\bibitem{angelini2010}
Angelini TE, Hannezo E, Trepat X, Fredberg JJ, Weitz DA.
\newblock Cell migration driven by cooperative substrate deformation patterns.
\newblock Physical Review Letters. 2010;104(16):168104.

\bibitem{oakes2018}
Oakes PW, Bidone TC, Beckham Y, Skeeters AV, Ramirez-San~Juan GR, Winter SP,
  et~al.
\newblock Lamellipodium is a myosin-independent mechanosensor.
\newblock Proceedings of the National Academy of Sciences.
  2018;115(11):2646--2651.

\bibitem{surface_evolver}
Brakke KA.
\newblock The surface evolver.
\newblock Experimental Mathematics. 1992;1(2):141--165.

\bibitem{miyata1995mechanical}
Miyata H, Yoshikawa H, Hakozaki H, Suzuki N, Furuno T, Ikegami A, et~al.
\newblock Mechanical measurements of single actomyosin motor force.
\newblock Biophysical Journal. 1995;68(4 Suppl):286S.

\bibitem{biron2005molecular}
Biron D, Alvarez-Lacalle E, Tlusty T, Moses E.
\newblock Molecular model of the contractile ring.
\newblock Physical Review Letters. 2005;95(9):098102.

\bibitem{yeung2005}
Yeung T, Georges PC, Flanagan LA, Marg B, Ortiz M, Funaki M, et~al.
\newblock Effects of substrate stiffness on cell morphology, cytoskeletal
  structure, and adhesion.
\newblock Cell motility and the Cytoskeleton. 2005;60(1):24--34.

\bibitem{sabass2008}
Sabass B, Gardel ML, Waterman CM, Schwarz US.
\newblock High resolution traction force microscopy based on experimental and
  computational advances.
\newblock Biophysical Journal. 2008;94(1):207--220.

\end{thebibliography}
\end{document}